\documentclass[10pt,twocolumn,twoside]{IEEEtran} 
\IEEEoverridecommandlockouts

\newcommand{\subparagraph}{}

\usepackage{graphicx,psfrag,dblfloatfix}
\usepackage{amsmath,amssymb,amsfonts,mathrsfs,theorem}
\usepackage{amsmath,amsfonts,amssymb,amscd,latexsym,enumerate}
\usepackage{graphicx,psfrag}
\usepackage{color,epstopdf}
\usepackage{bm}
\usepackage{enumerate} 
\usepackage{subfigure}
\usepackage{float}
\usepackage{bbold}
\usepackage{dsfont}
\usepackage{tikz}
\usepackage{verbatim}
\usepackage[ruled,longend]{algorithm2e}

\allowdisplaybreaks

\newtheorem{remark}{Remark}

\newtheorem{definition}{Definition}
\newtheorem{theorem}{Theorem}
\newtheorem{lemma}{Lemma}

\newtheorem{proposition}{Proposition}

\newcommand{\R}{\mathbb{R}}

\newcommand{\calW}{\mathcal{W}}    

\newcommand{\diag}{\operatorname{diag}} 

\def\qedp{\hspace*{\fill}~{\tiny $\blacksquare$}}
\def\be{\begin{equation}}
\def\ee{\end{equation}}
\def\ba{\begin{array}}
\def\ea{\end{array}}
\def\eqa{\begin{eqnarray}}
\def\eqe{\end{eqnarray}}

\SetVlineSkip{0pt}      

\definecolor{darkgreen}{rgb}{0.0, 0.55, 0.0}
\definecolor{amaranth}{rgb}{0.9, 0.17, 0.31}
\usepackage[prependcaption,colorinlistoftodos]{todonotes}

\usepackage{tikz}
\usepackage{tkz-berge}
\usetikzlibrary{positioning}
\usetikzlibrary{arrows,%
	petri,%
	topaths}%
\usetikzlibrary{decorations.markings}         
\tikzstyle{vertex}=[circle, shading = ball, ball color = white!100!white, minimum size = 15pt, draw, inner sep=0pt]  

\newcommand{\bbm}{\begin{bmatrix}}
	\newcommand{\ebm}{\end{bmatrix}}

\begin{document}

\title{
Bias estimation in sensor networks
}

\author{
Mingming Shi, Claudio De Persis, Pietro Tesi, Nima Monshizadeh
\thanks{
M. Shi, {C. De Persis}, P. Tesi  and N. Monshizadeh 
are with ENTEG, 
University of Groningen, 9747 AG Groningen, The Netherlands.
Email: {\tt\small M.Shi@rug.nl, n.monshizadeh@rug.nl, p.tesi@rug.nl, c.de.persis@rug.nl.}
P. Tesi is also with DINFO, University of Florence, 50139 Firenze, Italy 
E-mail: {\tt\small pietro.tesi@unifi.it}.}
}

\maketitle
\begin{abstract}
This paper investigates the problem of estimating  biases affecting relative state measurements in a sensor network. Each sensor measures the relative states of its neighbors and this measurement  is corrupted by a constant bias. We analyse under what conditions on the network topology and the maximum number of biased sensors the biases can be correctly estimated. We show that  for non-bipartite graphs  the biases can always be determined even when all the sensors are corrupted,
while
for bipartite graphs more than half of the sensors should be unbiased to ensure the correctness of the bias estimation. If the biases are heterogeneous, then the number of unbiased sensors can be reduced to two. Based on these conditions, we propose some algorithms to estimate the biases.
\end{abstract}

\section{Introduction}

%
The normal operation of many large scale systems relies on networks of sensors that provide information using for the monitoring and management of the system operating conditions \cite{Kekatos_DRPoweSysSE}-\nocite{Mitra2018DistributedOF,coddington2009rapid,Fagnani2014ADC,Fawzi2014SecureEA}\cite{Corts2009GlobalAR}. However, when measuring the  variables of interest, sensors may generate unreliable results due to the low quality of the hardware, environmental variations or adversary attacks. This introduces measurement errors, which can degrade the system performance and even lead to major disruptions \cite{Fawzi2014SecureEA}-\nocite{Corts2009GlobalAR,Mo2016OnTP,Zhao2016LocalizabilityAD,Li2013RobustDS,Marina2015ControllingRF,Fawzi2014SecureEA}\cite{Meng2016FormationCW}.

In this paper, we consider networks in which each sensor measures the difference between its state and that of its neighbors and aim to characterize the conditions under which the biases corrupting the measurements can  be estimated and provide methods for their estimation

The problem in this paper is broadly linked to others studied in the literature. 
Given erroneous relative measurements, providing precise estimates of the relative states can be considered as a complementary problem to the one of estimating biases. Many papers \cite {Zhao2016LocalizabilityAD,Barooah2007EstimationOG}-\nocite{Carron2014AnAC,Safavi2018DistributedLA,Shames2013AnalysisON,Wang2018DistributedNL,BoofNrobustdisest}\cite{Ravazzi2017DistributedEF} have provided methods for estimating the states of the sensors from noisy relative measurements by solving linear or nonlinear least square problems. 
These methods can not precisely estimate the state since the least square approach has no robustness to the measurement error and any error can make the estimation of the unknown deviate from the actual value \cite{Sharon2009}. 

The formulation of the problem considered in this paper covers the situation where the biases are constant but with  arbitrary magnitude, thus allowing for the presence of outliers.
Similar problems have been addressed recently in \cite{BoofNrobustdisest,Ravazzi2017DistributedEF}, where the focus is on the state estimation problem. 
However, neither one of the papers gives results on how the sparsity of the measurement errors affects the state estimation. On the other hand, computing biases from relative measurements received comparably less attention. 
The paper \cite{bolognani2010consensus} proposed  algorithms to estimate  sensor offsets in wireless sensor networks. These methods only partially compensate the offsets.
In problems that use the angle of arrival (AOA) measurements, if the local frame is unaligned with the global frame, then the unknown orientation of the local frame can be regarded as a bias. Ahn et al.~\cite{Lee2016DistributedFC,Oh2014FormationCA} use the consensus algorithm to estimate the orientation. However, similar to \cite{bolognani2010consensus}, the estimation error of their algorithms never vanish.

In this paper, we reduce the bias estimation problem to the  solution of linear equations (LEs). Several algorithms have been devoted to the distributed solution of LEs, with focus on asynchronous implementations \cite{lu2009distributed,Liu2018AsynchronousDA}, graph connectivity conditions \cite{Shi2017NetworkFT}, secure computing \cite{Shen2017ADS}, to name a few. However, in these algorithms, each node needs to find all the entries of the vector of the unknowns, which, if employed in our problem,  would  require the nodes to know the network size. Instead, we exploit a suitable sparsity condition on the biases to ensure they can be uniquely determined, which is an important problem in compressive sensing 
\cite{E.J.Candes_decoding}-\nocite{David_sparseidenti,DonohoDL_l0normopt,Bauhut_comsen_NUP,Rauhut_mathintr_CSl,Candes2008AnIT,ChenSS_BasisPursuit}\cite{Mota2012DistributedBP}, and is related to secure state estimation \cite{Fawzi2014SecureEA,hashimoto2018distributed}-\nocite{Chang2015SecureEB,Hu2018SecureSE}\cite{Shoukry2016ETCSO}.  

A related problem, which several papers have studied, is the one of achieving consensus or a prescribed formation in the presence of inconsistent or biased measurements. In \cite{Meng2016FormationCW}, the authors use estimators 
to counteract compass mismatches, 
while 
requiring each node to measure the relative positions of all the edges. 
The paper \cite{Marina2015ControllingRF} addresses the rigid formation control problem where the agents disagree on the prescribed inter-agent distances. For the problem considered in our paper, this method would require that for each pair of adjacent nodes, at least one of the nodes is bias-free. 
%
%
A similar set-up	 is also adopted in \cite{Liu2016RobustCO}.
For second-order consensus, \cite{Sukumar2018OnCO} proposes an adaptive compensator 
to prevent the state unboundedness caused by the biases.
The proposed compensator cannot make the system achieve exact consensus.

{\it Our contribution.} Given relative state measurements that are affected by biases, we 
find conditions under which the biases are identified so that the actual relative states can be exactly reconstructed.  
Similar to \cite{Kekatos_DRPoweSysSE,Zhao2016LocalizabilityAD,Carron2014AnAC,bolognani2010consensus,Lee2016DistributedFC,Oh2014FormationCA,Todescato-blockjacobi}, we assume that biased measurements can be exchanged among the neighboring nodes. 
Differently from \cite{BoofNrobustdisest,Sukumar2018OnCO}, we assume that 
each node has one sensor, hence the relative measurements taken by the node are affected by the same bias.
The form of the system of LEs to which we reduce the problem
is different from the one formulated in papers involving range or AOA measurements \cite{Meng2016FormationCW,bolognani2010consensus}-\nocite{Lee2016DistributedFC}\cite{Oh2014FormationCA}. In our problem (see Section \ref{sec:problem})
the biases affect the relative state measurements, whereas for problems involving range or AOA measurements the biases affect the absolute value of or the pointing of the vector of the relative measurements (distances or bearings). 
The LEs of the form considered in \cite{Meng2016FormationCW,bolognani2010consensus}-\nocite{Lee2016DistributedFC}\cite{Oh2014FormationCA} 
also appears in papers that studied problems  of sensor synchronization \cite{Giridhar2006DistributedCS} and multi-agent fault estimation \cite{hashimoto2018distributed}.  

We provide conditions under which the biases are uniquely determined from the proposed system of LEs. Our results answer the question: ``what is the maximum number of sensor biases  that can be estimated  from erroneous 
relative state measurements?" For non-bipartite graphs, the answer is ``all the nodes" and we provide a distributed algorithm to estimate the biases. In the algorithm, each sensor only needs to estimate its own bias, leading to a reduction of the computational resources and memory sizes required at each node, a solution that  is different from those in \cite{lu2009distributed}-\nocite{Liu2018AsynchronousDA,Shi2017NetworkFT}\cite{Shen2017ADS}. 

For bipartite graphs, similar to secure state estimation problems \cite{Fawzi2014SecureEA,hashimoto2018distributed}-\nocite{Chang2015SecureEB,Hu2018SecureSE}\cite{Shoukry2016ETCSO}, we show that the biases can be correctly computed when less than half of the sensors is biased. Furthermore, 
we prove that the maximum number of biased sensors can be increased if the biases are heterogenous. This reduces the number of unbiased sensors to only two and  
improves the results in secure state estimation. We provide two 
algorithms to compute the biases. By exploiting the heterogeneous assumption and a coordinator to coordinate the sensors, the first algorithm we propose computes the biases in a finite number of steps. To remove the coordinator and make the estimation fully distributed, in the second algorithm we solve a relaxed 
$\ell_1$-norm optimization problem as in \cite{hashimoto2018distributed,Hu2018SecureSE}. We show an interesting result that the actual  vector of biases is the unique solution of the $\ell_1$-norm optimization problem if less than half of the sensors are biased, which does not worsen the bound on the sparsity condition of the biases for the non-relaxed problem. 

We also apply the bias estimation algorithms to a consensus problem. Different from \cite{Sukumar2018OnCO}, we can prove that the system achieves exact consensus. Our algorithms do not require each node to measure the relative states of all the edges, in contrast to \cite{Meng2016FormationCW}.

The rest of the paper is organized as follows. In Section \ref{sec:Pre}, we introduce the notation, some general notions about graphs and and few specialized results on bipartite graphs. We formulate the problem and provide a useful lemma in Section \ref{sec:problem}. Section \ref{sec:nonbipttgraph} deals with  the bias estimation algorithm for non-bipartite 
graphs.
In Section \ref{sec:bipttgraph}, we introduce the sparsity condition on biases that ensures the correctness of the bias estimation, we provide two bias estimation algorithms and show consensus using one of the proposed algorithms. Section \ref{sec:examp} presents numerical experiments 
to validate the theoretical findings. 

\section
{Preliminaries
} \label{sec:Pre}
\subsection{
	Notation}
For a vector $z\in \R^p$, $\mathrm{diag}\{z\}$ represents the diagonal matrix with the $i$th diagonal entry equal to the $i$th element of $z$. We denote by $\mathcal{S}_z$ the support of $z$, which is the set of indices that correspond to the nonzero entries of $z$, and by $\|z\|_0$ the $0$-norm of $z$, which is the number of elements in $\mathcal{S}_z$. We let $\mathbb 1_{m}$ and $\mathbb 0_m$ denote the $m$-dimensional vectors with all elements equal to $1$ and $0$, respectively. Given a matrix $A$, $A_i$ represents its $i$th row and $a_{ij}$ represents its element in the $i$th row and  $j$th column. The cardinality of a set $S$ is denoted by $|S|$. For two sets $S$ and $M$, we let $S\setminus M=\{x\in S \mid x\notin M\}$ represent the complement  of $M$ in $S$.

\subsection{{Graph-theoretic notions}}\label{sec:graph.not}
For a network with $n$ nodes, let its topology be represented by an undirected and connected graph $G=\{V,E\}$, with $V=\{1,2,...,n\}$ being the set of nodes and $E\subseteq V\times V$ be the set of edges, where $\{i, j\}\in E$, or equivalently, node $i$ is a neighbour of node $j$,  means that node $i$ can receive information from node $j$ and vice versa.  We denote the  set of neighbors of node $i$ by $\mathcal{N}_i$,  and let $d_i=|\mathcal{N}_i|$.

The adjacency matrix $A$ of $G$ is defined as $a_{ij}=1$ if node $j$ is the neighbor of node $i$ and $a_{ij}=0$ otherwise. 
For an undirected graph $G$, we can assign arbitrary orientations to the edges such that each edge $\{i,j\}\in E$ has a head and a tail. The edge-node incidence matrix $B\in \R^{m\times n}$ of $G$, with $m=|E|$, is defined as $b_{ij}=1$ if $j$ is the head node of the edge $i\in E$ and $b_{ij}=-1$ if $j$ is the tail node. The Laplacian matrix $L$ of $G$ is an $n\times n$ matrix given by $l_{ij}=-a_{ij}$ for $j\neq i$ and $l_{ii}=\sum_{j\in \mathcal{N}_i}a_{ij}=d_i$. Since $G$ is undirected, it is well-known that $L=B^\top B$. 
The incidence matrix can be decomposed as the head incidence matrix $B_{+}\in \R_{m\times n}$ and the tail incidence matrix $B_{-}\in \R_{m\times n}$, which are given by
\begin{eqnarray}
b_{+,ij}&=&\left\{\begin{array}{cc}
1, & \mathrm{if\ node }\ j \ \mathrm{is\ the\ head}\\
0, & \mathrm{otherwise}\qquad\qquad\quad
\end{array}
\right.\nonumber\\
b_{-,ij}&=&\left\{\begin{array}{cc}
-1, & \mathrm{if\ node }\ j \ \mathrm{is\ the\ tail}\\
\,0, & \mathrm{otherwise}\qquad\qquad\
\end{array}
\right.\nonumber
\end{eqnarray}
We also let $R$ denote the signless edge-node incidence matrix with $r_{ij}=|b_{ij}|$. It is easy to verify that $B=B_{+}+B_{-}$ and $R=B_{+}-B_{-}$. Let $d=[d_1\; d_2\; ...\;d_n]^\top$ and $D=\diag\{d\}$. The matrix $A+D$ is called the signless Laplacian matrix. 
When $G$ is undirected, $A+D=R^\top R$. Hence, $A+D$ is positive semi-definite and all its eigenvalues  $\mu_1\le \mu_2\le\cdots\le \mu_n$ are real and nonnegative.

A path $\mathcal{P}_{ij}$ from node $i$ to node $j$ is a sequence of nodes and edges such that each successive pair of nodes in the sequence is adjacent. 
The length of a path is the number of edges in the path. The distance between node $i$ and $j$ is the length of the shortest path from $i$ to $j$. 
We denote 
by $D_G$ the diameter of $G$, which is the maximum distance between any two nodes. 

\subsection{Bipartite graphs}
A graph $G$ is bipartite if the vertex set $V$ can be partitioned into two sets $V_+$ and $V_-$ in such a way that no two vertices from the same set are adjacent. The sets $V_+$ and $V_-$ are called the colour classes of $G$ and $(V_+,V_-)$ is a bipartition of $G$. For a bipartite graph, the following result holds:
\begin{theorem}\label{thm:bipp_oddcycle}
{\rm \cite{Asratian_Armen_S}}	A graph $G$ is bipartite if and only if $G$ has no cycle of odd length.
\end{theorem}

An algebraic characterization of bipartite graphs is provided next.
\begin{lemma}\label{lem:R_nonbiptt}
An undirected and connected graph $G$ is bipartite if and only if the signless incidence matrix $R$ does not have full column rank. Moreover, if $G$ is bipartite, then any $n-1$ columns of $R$ are linearly independent.
\end{lemma}
%
\emph{Proof.} To prove the first part, suppose that $Rv = 0$ for some nonzero vector $v\in \R^{n}$. It is easy to see that $|v_i|=|v_j|=a$ for every $i,j \in V$, where $a>0$. In fact, consider any path connecting nodes $i$ and $j$. For every pair $(r,s)$ of adjacent nodes in this path we must have $v_r = -v_s$ otherwise $Rv \neq 0$. Since the graph is connected and since $v$ must be nonzero, we obtain the claim. Thus there exists a bipartition $(V_+,V_-)$ of $G$, where the nodes corresponding to the entries of $v$ with value $a$ and $-a$ are assigned to $V_+$ and $V_-$, respectively. Conversely if $G$ is bipartite, there exists a bipartition $(V_+,V_-)$ of $G$. By letting the elements of $v$ corresponding to $V_+$ and $V_-$ be $a$ and $-a$, respectively, with $a\ne 0$, we have $Rv=0$, which shows that $R$ does not have full column rank.

For the second part, we prove it by contradiction. Suppose there exist some dependent columns of $R$ and let the index set of these columns be $S\subset V$, with $|S|\le n-1$, then there should exist a nonzero vector $v\in \R^{|S|}$ such that $R_Sv=0$ where $R_S$ is the matrix whose columns are those indexed by $S$. The latter implies the existence of a nonzero vector $\tilde v$, whose nonzero entries are given by $v$,  and 
satisfies $R\tilde v=0$. 
However, from the proof of the first part, the absolute values of all the elements of $\tilde v$ should be equal to each other. Hence, $v$ must be the zero vector, which is a contradiction.\qedp

%
%

The if and only if part of the statement above is also provided in \cite[Lemma 2.17]{bapat2010graphs}. We provide the proof here, since it is used in proving the second part of the statement as well as in other parts of the paper.

For  later use, by the proof of Lemma \ref{thm:bipp_oddcycle}, we note that 
\begin{equation}\label{eq:KerR}
Rv= \mathbb{0}_n  \Longleftrightarrow  \exists a\in \R  {\;\rm s.t.\;} v_i= 
\begin{cases}  
a &  \; i\in V_+ \\
-a & \; i\in V_-
\end{cases}
\end{equation}
 for a bipartite graph with bipartition $(V_+, V_-)$.
\begin{lemma}\label{lem:bipp_sglessLpsp}
{\rm \cite{Cvetkovic_Dragos}}
	The smallest eigenvalue of the signless Laplacian matrix $A+D$ of an undirected and connected graph is equal to zero if and only if the graph is bipartite. In case the graph is bipartite, zero is a simple eigenvalue.
\end{lemma}
 
\subsection{Compressed sensing}\label{subsec:decoding}
In the field of compressed sensing or sparse signal recovery, one of the most important problems is how to find the sparsest solution from the number-deficient measurements. Formally, consider the following linear equation
\begin{eqnarray}\label{eq:sps_compsensing}
y=Fx
\end{eqnarray}
where $x\in \R^n$ is the vector of unknown variables, $y\in \R^{p}$ is the vector of  known values, and $F\in \R^{p\times n}$ is a matrix defining the linear relation from $x$ to $y$. 
It is assumed that $p<n$, thus equation \eqref{eq:sps_compsensing} is under-determined. It is then of interest to find solutions $x$ such that  $\|x\|_0\ll n$, and in particular to seek for the sparsest solution of \eqref{eq:sps_compsensing}. {Let us define the set of $k$-sparse vectors as  
	\be
	\calW_k:=\{x\in \R^n \mid \|x\|_0 \leq k\}.
	\ee
	}
The following result provides a sufficient condition under which the solution of \eqref{eq:sps_compsensing} can be uniquely determined.
\begin{lemma}\label{lem:compsensing_unique} 
	Given an integer $s\ge 0$, let $2s\le p$, and assume that any matrix made of $2s$ columns of $F$ is full {column} rank. 
	If $x{\in \mathcal{W}_s}$ is a solution of \eqref{eq:sps_compsensing}, then {there exists no other solution of \eqref{eq:sps_compsensing} in $\mathcal{W}_s$. 
	}
\end{lemma}

\begin{remark}
\rm{ Under the assumptions of the lemma, the solution $x\in \mathcal{W}_s$ of \eqref{eq:sps_compsensing} is also the solution to 
\be\label{eq:min.support}
\ba{rl}
\displaystyle\min_{\overline x\in \mathbb{R}^n}  & \|\overline x\|_0\\
{\rm s.t.}  & y=F\overline x,
\ea
\ee
that is, the sparsest solution to
\eqref{eq:sps_compsensing}. The proof of Lemma \ref{lem:compsensing_unique} descends from \cite[Lemma 1]{David_sparseidenti}.} \qedp
\end{remark}
However, solving $x$ from \eqref{eq:sps_compsensing} under the assumption that $\|x\|_0\le s$ is cumbersome when $s$ is not small, as it requires to combinatorially search {for $s$ columns of $F$ whose span contains $y$. }
A typical way to avoid this exhaustive search is to change the problem into the following $\ell_1$-norm optimization problem
\begin{eqnarray}\label{eq:l_1min}
\min_{\overline x\in \mathbb{R}^n} && \|\overline x\|_1\\
\mathrm{s.t.} && y= F\overline x\nonumber
\end{eqnarray}
where $y$ is the vector of known values in \eqref{eq:sps_compsensing} and the objective function and the constraint are both convex. 
Problem \eqref{eq:l_1min} can be solved by linear programming \cite{Bauhut_comsen_NUP}. 
The $\ell_1$-norm minimization may return a solution {$\overline x_*$ different from 
the solution $x$ of \eqref{eq:sps_compsensing}.  
}
The following definition and result characterize the relation between the matrix $F$, {the equation \eqref{eq:sps_compsensing}} and 
the $\ell_1$-norm minimization problem.
\begin{definition}[Nullspace Property]\label{def:nup}
	A matrix $F\in \R^{p\times n} $ is said to satisfy the nullspace property of order $s$, with $s$ 
	being a {positive}  
	integer, 
	if for any set $S\subset V=\{1,2,...,n\}$ with $|S|\le s$ and {any nonzero vector} $v$ in the null space of $F$, the  condition below holds
	\begin{eqnarray}\label{eq:nup}
	\|v_{S} \|_1 < \|v_{S_c}\|_1, 
	\end{eqnarray} 
	where $v_S\in \mathbb{R}^{|S|}$ and $v_{S_c}\in \mathbb{R}^{|S_c|}$ are subvectors of $v$ whose elements are indexed by $S$ and $S_c$, respectively, and $S_c=V\setminus S$.
\end{definition}
{The null space property is usually difficult to verify and a more restrictive but more conveniently checkable condition known as
restricted isometry property is considered \cite[p.~8]{Bauhut_comsen_NUP}. Yet, in the special cases that are of interest to us the null space property can be easily confirmed (cf., Theorem \ref{thm:bias_ell_1min}), and we will persist with it in the sequel. }
\begin{theorem}
	\label{thm:nup_recovery} {\rm \cite[Theorem 2.3]{Bauhut_comsen_NUP}}
	Every vector 
	$x\in \mathcal{W}_s$
	is the unique solution of the $\ell_1$-norm minimization problem \eqref{eq:l_1min}, with $y=Fx$, if and only if $F$ satisfies the null space property of order $s$.
\end{theorem}
We highlight the role of this theorem explictly in connection with the equation \eqref{eq:sps_compsensing}. For a given $y\in \mathbb{R}^p$, let $x\in \mathbb{R}^n$ be a solution of
\eqref{eq:sps_compsensing}. Assume that $\|x\|_0\le s$ and $F$ satisfies the null space property of order $s$, with $0<s<n$. By Theorem  \ref{thm:nup_recovery}, $x$ is the unique solution of \eqref{eq:l_1min}, with $y=Fx$. Stated directly, there exists a unique solution $\overline x_*$ of \eqref{eq:l_1min}, with $y=Fx$, and it satisfies $\overline x_*=x$. Hence, under the given condition of $s$-sparsity of the vector $x$ solution of \eqref{eq:sps_compsensing} and the null space property of order $s$ of the matrix $F$, solving the optimization problem \eqref{eq:l_1min}, with $y=Fx$, univocally returns $x$.  


%
%

\section{Problem formulation -- biases estimation in sensor networks}\label{sec:problem}
We consider a sensor network where each sensor is identified with a node in a graph $G=(V,E)$ with $V$ the set of nodes, $|V|=n\ge 2$ and $E$ the set of edges. {Throughout the paper, we assume that $G$ is connected and undirected.} A state variable $x_i\in \mathbb{R}$ is associated to each node $i\in V$.  Each sensor $i\in V$ can measure the relative information  $x_j-x_i$ for all $j\in \mathcal{N}_i$.


We are interested in a scenario where the measurements taken by the sensor network may be subject to constant biases. As a result of the bias, the relative information read by the sensor $i$, 
will be modified as 
\begin{equation}\label{eq:biased_meas}
z_{ij}=x_{j}-x_i+w_i, \quad \forall j\in \mathcal{N}_i,
\end{equation}
where $w_{i}\in \R$ is an unknown constant term accounted for the bias of sensor $i$. In case a sensor is bias free we set $w_i=0$.

The presence of biases deteriorate the performance of the network, and may even raise stability issues.  
Thus it is of interest to estimate the biases, and possibly counteract their effect in the network. 

To formulate the problem, we first rearrange the equalities in  \eqref{eq:biased_meas} in a suitable vector form. 
After assigning arbitrary orientation to $G$,  we collect in the vector $\zeta\in \mathbb{R}^m$ all the measurements $z_{ij}$ for which node $i\in V$ is the head of the edge ${i,j}\in E$, which gives $\zeta= -Bx +B_+ w$, with $B_+$ denoting the head incidence matrix. Similarly, we collect in the vector $\eta\in \mathbb{R}^m$ all the measurements $z_{ij}$ for which node $i\in V$ is the tail of the edge ${i,j}\in E$, and obtain $\eta= Bx -B_- w$, where $B_-$ is the tail incidence matrix. Hence, 
\begin{eqnarray}\label{eq:decoding}
z:=
\begin{bmatrix}
\zeta\\
\eta
\end{bmatrix}
=\left[\begin{array}{r}
-B\\
B
\end{array}\right] x
+\left[\begin{array}{r}
B_{+}
\\
-B_{-}
\end{array}\right]w
\end{eqnarray} 
Note that, by construction,  we have $z\in {\rm im}(\mathcal{B})$, where
\[
 \mathcal{B}:= \left[\ba{cc} -B & B_+\\ B & -B_- \ea\right]
\] 
and $\rm{im}(\cdot)$ denotes the column span of a matrix. 

For a given measurement $z$, we are interested in finding the bias vector $w$ in a set $\mathcal{W}\subseteq \R^{n}$ of  admissible biases, which is defined more precisely later. To avoid ambiguity, we first introduce the definition of a solution of \eqref{eq:decoding} with respect to $w$.
\begin{definition}[Solution of \eqref{eq:decoding} in $\mathcal{W}$]\label{def:sol.in.W}
	Given $z\in \rm{im} \left(\mathcal{B}\right)$ and a set $\mathcal{W}\subseteq \R^{n}$ of admissible biases, the vector $\overline w\in \mathcal{W}$ solves \eqref{eq:decoding} if there exists $\overline x\in \R^n$ such that \eqref{eq:decoding} %
	is satisfied with  $( x, w)=(\overline x, \overline w)$.  In this case, we say {$\overline w$ solves \eqref{eq:decoding} in $\mathcal{W}$, or $\overline w$ is a solution of \eqref{eq:decoding} in $\mathcal{W}$.}
\end{definition}
	
The uniqueness of the solution of \eqref{eq:decoding} is defined below:

\begin{definition}[Unique solution of \eqref{eq:decoding} in $\mathcal{W}$]
	A solution $\overline w$ of \eqref{eq:decoding} in $\mathcal{W}$ is unique if
there exists no  vector {$\overline w^{\,\prime}$,} with 
$\overline w^{\,\prime} \neq \overline w$,  {which is a solution} of \eqref{eq:decoding} in $\mathcal{W}$.
{In this case, we say $\overline w$ uniquely solves \eqref{eq:decoding} in $\mathcal{W}$.}
\end{definition}

We then formulate the problem which is of interest in this paper. 
	
	\smallskip	
{\it Problem formulation. } 
 Given the vector of biased measurements $z\in \rm{im} \left(\mathcal{B}\right)$ and a set $\mathcal{W}\subseteq \R^{n}$ of admissible biases, find conditions under which the vector of actual sensor biases $w$ is the unique solution of \eqref{eq:decoding} in $\mathcal{W}$, and design 
 algorithms for estimating it. 
%

\smallskip
Note that $\mathcal{W}$ should always contain the bias vector $w$ and by construction at least one solution to \eqref{eq:decoding} exists. Determining conditions under which the solution to  \eqref{eq:decoding} is unique implies we can correctly estimate the vector of actual biases affecting the measurements. To prove the uniqueness of the solution of  \eqref{eq:decoding} we will rely on a reduced form of \eqref{eq:decoding} provided in the following result:

\begin{lemma}\label{lem:ifpart}
Consider the vector of biased measurements $z\in \rm{im} \left(\mathcal{B}\right)$ and {a} set $\mathcal{W}\subseteq \R^{n}$ of admissible biases. Consider the equality 
\begin{eqnarray}\label{eq:Rw}
Rw=\tilde z
\end{eqnarray}
where $R=B_{+}-B_{-}$ is the signless edge-node incidence matrix, $\tilde z=Fz$, and $F=[I_m\; I_m]$ is the left annihilator of the matrix $[-B^\top\; B^\top]^\top$. Then the following two statements hold:
\begin{enumerate}[(i)]
	\item The vector $w$ is a solution of \eqref{eq:decoding} in $\mathcal{W}$ if and only if $w\in \mathcal{W}$ is a solution of \eqref{eq:Rw}.
	\item The vector $w$ is the unique solution of \eqref{eq:decoding} in $\mathcal{W}$ if and only if $w$ is the unique solution of \eqref{eq:Rw} in $\mathcal{W}$.
\end{enumerate}
\end{lemma}

\emph{Proof.} $(i)$. {(Only if)} If $w$ is a solution of \eqref{eq:decoding} in $\mathcal{W}$, then pre-multiplying \eqref{eq:decoding} by $F$ leads to $\tilde z=Rw$. Hence $w\in \mathcal{W}$ is also a solution of \eqref{eq:Rw}. 

{(If)} Since $w$ is a solution of \eqref{eq:Rw} in $\mathcal{W}$, then $\zeta+\eta=Rw$, with $w\in \mathcal{W}$. Since $z \in {\rm im}(\mathcal{B})$, there should exist a vector $x^{\prime}\in\R^n$ and $ w^{\prime}\in \R^n$ such that 
\begin{eqnarray}\label{miserve}
	z=
	\left[\begin{array}{r}
	-B\\
		B
	\end{array}\right] x'
	+\left[\begin{array}{r}
		B_{+}
		\\
		-B_{-}
	\end{array}\right]w'
\end{eqnarray} 
Pre-multiplying the equality above {by $F$} leads to $\tilde z=Rw'$. Combining this with ${\tilde z}=Rw$, we have $R(w'-w)=\mathbb 0_m$. We continue the proof considering the following two distinct cases.

{\it Case 1. $G$ is not bipartite.} Since $G$ is not bipartite, by Lemma 1 the matrix $R$ is full-column rank, which implies $w'=w\in \mathcal{W}$. Hence $w\in \mathcal{W}$ is a solution of \eqref{eq:decoding}.

{\it Case 2. $G$ is bipartite.} Since $G$ is bipartite, there should exist a bipartition $V=\{V_+,V_-\}$.  Let $|V_+|=p$, label the nodes in $V$ such that $V_+=\{1,2,...,p\}$, $V_-=\{p+1,...,n\}$ and define the orientations of the edges in such a way that the head node of each edge in $E$ belongs to $V_+$. 
{Bearing in mind the identity $R(w'-w)=\mathbb 0_m$ above, and noting \eqref{eq:KerR} we have}
\begin{equation}\label{eq:f}
w'=w+ fa , \quad f= \bbm  \mathbb{1}_p \\ - \mathbb{1}_{n-p} \ebm,
\end{equation}
for some $a\in \R$.
Substituting this back to \eqref{miserve} yields
\begin{equation}\label{eq:pr-lem4-1}
z=	\left[\begin{array}{r}
- B\\
B
\end{array}\right] x'
+\left[\begin{array}{r}
B_{+}
\\
-B_{-}
\end{array}\right]w 
+\left[\begin{array}{r}
B_{+}
\\
-B_{-}
\end{array}\right]fa.
\end{equation}
{To prove that $w$ is a solution of \eqref{eq:decoding} in $\mathcal{W}$}, 
in view of Definition \ref{def:sol.in.W},
we need to show that
\begin{equation*}
z- \bbm 
B_+\\
-B_-
\ebm w \in
\mathrm{im} \bbm -B \\  B \ebm,
\end{equation*}
which, by \eqref{eq:pr-lem4-1}, reduces to
\begin{equation}\label{eq:pr-lem4-2}
\left[\begin{array}{r}
B_{+}
\\
-B_{-}
\end{array}\right]f
\in
\mathrm{im} \bbm  -B \\  B \ebm.
\end{equation}
{Let $B_+$ and $B_-$ be decomposed as
\[
B_+=\bbm \tilde B_+ & \mathbb{0}_{m\times (n-p)} \ebm, \quad  B_-=\bbm \mathbb{0}_{m\times p} & \tilde B_{-}\ebm
\]
for some matrices $\tilde B_+$ and $\tilde B_-$. Then \eqref{eq:pr-lem4-2} can be written as
}
\[
\left[\begin{array}{rr}
\tilde{B}_{+} & \mathbb{0}_{m\times (n-p)}\\
\mathbb{0}_{m\times p} & -\tilde{B}_-
\end{array}\right]f
\in
\mathrm{im} \bbm  
-\tilde{B}_+ & -\tilde{B}_-\\ \tilde{B}_+ & \tilde{B}_-
\ebm
\]
{where we have used the fact that $B=B_++B_-$.}
Noting that $\tilde{B}_+ \mathbb{1}_{p}=-\tilde{B}_- \mathbb{1}_{n-p}$,
 it is easy to verify that the above {relationship} is satisfied since
\[
\left[\begin{array}{rr}
\tilde{B}_{+} & \mathbb{0}_{m\times (n-p)}\\
\mathbb{0}_{m\times p} & {-} \tilde{B}_-
\end{array}\right]f
=
 \bbm  
-\tilde{B}_+ & -\tilde{B}_-\\ \tilde{B}_+ & \tilde{B}_-
\ebm
  \bbm  \mathbb{0}_{p} \\ \mathbb{1}_{n-p} \ebm. \]
This completes the proof of part (i).

$(ii)$. We only prove the ``if" part since the {converse implication} can be shown {similarly}. Assume $w$ is a unique solution of \eqref{eq:Rw} in $\mathcal{W}$, then by (i), we have $w$ is also a solution of \eqref{eq:decoding} in $\mathcal{W}$. Now if there exists another vector $w'\in \mathcal{W}$, with $w'\ne w$ is a solution of \eqref{eq:decoding}, it should also be a solution of \eqref{eq:Rw} by the first statement. This contradicts the uniqueness assumption. \qedp

{The result of Lemma \ref{lem:ifpart} will be used in some of the derivations of the main results in the sequel.}

{To study the conditions guaranteeing the uniqueness of the solution of \eqref{eq:decoding} in $\mathcal{W}$,
we differentiate between bipartite and not bipartite graphs.}


\section{Non-bipartite graphs}\label{sec:nonbipttgraph}
In this section, we present the results for the case when the measurement graph $G$ is not bipartite. 
%
\subsection{Condition for correct bias estimation}
The following result shows that  $w$ can  be determined uniquely from \eqref{eq:decoding} if the graph is not bipartite.
\begin{theorem}\label{thm:suff-condt}
	Consider a graph $G$, 
let 
{$z\in \mathrm{im} (\mathcal{B})$} be the vector of biased measurements, and $\mathcal{W}=\R^n$ be the set of admissible biases. 
Then $w$ is the unique solution of \eqref{eq:decoding} in $\mathcal{W}=\R^n$ if and only if $G$ is not bipartite.
\end{theorem}
\emph{Proof}.
In view of Lemma \ref{lem:ifpart}, we need to show  that the bias vector $w$ is the unique solution of \eqref{eq:Rw}
 if and only if $G$ is not bipartite. This holds since, by Lemma \ref{lem:R_nonbiptt}, the matrix $R$ has full column rank if and only if $G$ is not bipartite. \qedp


%

\subsection{Distributed bias estimation}\label{subsec:distest_nonbip}

In this section we propose a distributed algorithm to estimate the biases. We assume the existence of a communication network, modeled by an undirected and connected graph $G_c=(V_c, E_c)$, through which the nodes can communicate with each other without any imperfection. We let  $V_c=V$ and $E_c=E$.   

We assign to each node a bias estimation variable $\hat w_i$ of the bias $w_i$ affecting its sensor.  For each node $i\in V$, we let the estimation variable evolve as follows:
\begin{equation}\label{eq:biasest}
\dot {\hat w}_i=\sum_{j\in \mathcal{N}_i}(z_{ij}+z_{ji}-\hat w_i-\hat w_j)
\end{equation}
Node $i$ uses the biased measurements $z_{ij}$ and $z_{ji}$, and the bias estimates $\hat w_i$ and $\hat w_j$. Note that the values of $z_{ji}$ and $\hat w_j$ are communicated to node $i$ via the link $\{j, i\}$.

The following result shows exponential convergence of the estimates to the actual biases. 
\begin{proposition}\label{thm:biasest}
	The estimate vector $\hat w$ generated by \eqref{eq:biasest} converges exponentially fast to the vector $w$ of the actual biases  {if} the measurement graph G is not bipartite.
\end{proposition}
\emph{Proof.} Denote the estimation error for the bias $w_i$ as $e_i={\hat w_i}-w_i$. 
From \eqref{eq:biasest}, we have 
\begin{eqnarray}\label{eq:esterr}
\dot e_i&=&\dot {\hat w}_i-\dot w_i\nonumber\\
&=& \sum_{j\in \mathcal{N}_i}(z_{ij}+z_{ji}-\hat w_i-\hat w_j)\nonumber\\
&=& \sum_{j\in \mathcal{N}_i}(x_j-x_i+w_i+x_i-x_j+w_j-\hat w_i-\hat w_j)\nonumber\\
&=& -\sum_{j\in \mathcal{N}_i}( e_i+e_j)
\end{eqnarray}
which in a matrix form can be expressed as 
\begin{equation}\label{eq:esterrdyn}
\dot e= -(A+D)e 
\end{equation}
By Lemma \ref{lem:bipp_sglessLpsp}, the matrix $-(A+D)$ is Hurwitz if and only if $G$ is not bipartite. The exponential convergence of the estimation error $e$ then follows immediately. \qedp

\smallskip

An alternative way to solve for $w$ in \eqref{eq:Rw} is to use the block partition method of 
\cite{IvanoNotarnicola,Todescato-blockjacobi,BoofNrobustdisest}. When applied to the problem under investigation in this paper, the method requires each node to estimate not only its own bias but also those of its neighbors.  In contrast, the estimation algorithm \eqref{eq:biasest} only requires each node to store and transmit it own estimate, hence it reduces the memory space and communication burden. 

\subsection{An example of use: rejecting biases in a consensus network}\label{subsec:bias.rejection}
In this subsection, we investigate the possibility of removing  the effect of relative state measurement biases from a consensus algorithm. By exploiting the bias estimation method provided in the previous subsection, we
devise a compensator that asymptotically rejects the biases. To this end, let 
\begin{eqnarray}\label{eq:dyn_bias}
\dot x_i&=&\sum_{j\in \mathcal{N}_{i}}z_{ij}+u_i^c \nonumber \\
&=&\sum_{j\in \mathcal{N}_{i}} (x_j-x_i)+d_iw_{i}+u_i^c, \quad \forall i\in V
\end{eqnarray}
where $u_i^c$ is an additional control input available to the designer.  
Note that without a proper compensation, i.e., $u_i^c=0$, solutions of \eqref{eq:dyn_bias} can be {unbounded.}  
 Let $u_i^c$ be given by
\begin{eqnarray}\label{eq:compensator}
u_i^c=-d_i\hat w_i, \quad \forall i\in V
\end{eqnarray}
where $\hat w_i$ is given by  \eqref{eq:biasest}. 
This  results in the closed-loop dynamics
\begin{eqnarray}\label{eq:biascomp}
\dot x_{i}&=&\sum_{j\in \mathcal{N}_i}y_{ij}-u_i^c\nonumber\\
&=&\sum_{j\in \mathcal{N}_i}(x_j-x_i+w_i-\hat w_i)\nonumber\\
&=& \sum_{j\in \mathcal{N}_i}(x_j-x_i-e_i)
\end{eqnarray}
which can be written compactly as 
\begin{eqnarray}\label{eq:dynbiaserr}
\dot x&=& -Lx-De.
\end{eqnarray}

In case of a non bipartite graph,  the vector of biases can be asymptotically rejected and consensus can be achieved:
\begin{proposition}\label{thm:consensus}
{Let $G$ be a non-bipartite graph}.  
Then, solutions $(e, x)$ of \eqref{eq:esterrdyn}, \eqref{eq:dynbiaserr}, exponentially  converge to the point $(e^*, x^*)$, 
where $x^*\in \mathrm{im}(\mathbb{1}_n)$ and $e^*=0$. 
If $\hat w$ is initialized at zero, equivalently $e(0)=-w$, then we have
\begin{equation}\label{eq:xi*}
{x_i^*}=\frac{\mathbb{1}_n^\top}{n}D(A+D)^{-1}w+\frac{\mathbb{1}_n^\top  x(0)}{n}. 
\end{equation}
for each $i\in V$.
\end{proposition}

\emph{Proof}. Equation \eqref{eq:dynbiaserr} can be seen as the conventional consensus dynamics driven by the bias estimation error. Let $0=\lambda_1<\lambda_2\le\lambda_3\le\cdots\le\lambda_n$ be the eigenvalues of $L$ along with the basis of orthonormal eigenvectors 
$\{\frac{\mathbb{1}_n}{\sqrt{n}},v_2,\cdots,v_n\}$. Define $\Lambda=\diag{[\lambda_1,\cdots,\lambda_n]}$, $U=[\frac{\mathbb{1}_n}{\sqrt{n}} \; U_2]$ with $U_2=[v_2\;\cdots\;v_n]$ and apply the state transformation $z=U^\top  x$.
In the new coordinates, we have
\begin{eqnarray}
\dot z=-\Lambda z-U^\top De
\end{eqnarray} 
where $z_1$ is the solution of 
\begin{equation}\label{eq:dotz1}
\dot z_1=-\frac{\mathbb{1}_n^\top D}{\sqrt{n}}e
\end{equation} 
and {$z_{[2:n]}:=\bbm z_2 \;  \dots \; z_n\ebm^\top$} follows
\begin{equation}\label{eq:dotz2N}
\dot z_{[2:n]}=-\overline \Lambda z_{[2:n]}-U_2^\top De
\end{equation}
with $\overline \Lambda=\diag{[\lambda_2,\cdots,\lambda_n]}$. 
By Proposition \ref{thm:biasest}, if $G$ is not bipartite then the estimation errors satisfy
\begin{equation}
\label{eq:biasesterr}
e(t)=e^{-(A+D)t}e(0),
\end{equation}
from which we have 
\begin{equation}
z_1(t)=-\frac{\mathbb{1}_n^\top }{\sqrt{n}}D(A+D)^{-1}(1-e^{-(A+D)t})e(0)+z_1(0)
\end{equation}
which implies $$\lim_{t\to +\infty} z_1(t)=-\frac{\mathbb{1}_n^\top }{\sqrt{n}}D(A+D)^{-1}e(0)+z_1(0).$$ Since $\overline \Lambda>0$,  the vector $z_{2:n}(t)$ converges to zero exponentially fast. 
{Hence, we find that $x$ exponentially converges to $c\mathbb{1}_n$ for some $c\in \R$.
It is easy to see that
\[
c= \frac{1}{\sqrt{n}}  \lim_{t\rightarrow +\infty}  z_1(t).
\]
If  $e(0)=-w$, then $c=x_i^*$ given by \eqref{eq:xi*}, for each $i\in V$, which completes the proof.} 
\qedp

\smallskip

Although the system with bias compensation achieves consensus, the exact consensus value  to which the agents converge is not predictable since it depends both on the initial state and the bias of the sensors. For those problems where it is of primary interest to converge to the average consensus, alternatively  one can first run the algorithm \eqref{eq:biasest} over a sufficiently large time horizon to obtain a sufficiently accurate estimate of the biases, and then directly remove the biases from the measurements used in the consensus algorithm. 

\section{Bipartite graphs}\label{sec:bipttgraph}

In this section, we consider the case where the measurement graph $G$ is bipartite.

\subsection{Conditions for bias estimation}
For bipartite graphs, the following result gives a general condition that ensures that the vector of biases can be correctly estimated from the measurement \eqref{eq:decoding}.
\begin{theorem}\label{thm:suff-condt_biptt}
Consider a bipartite graph $G$,
{if a vector $w$ solves \eqref{eq:decoding} in $\mathcal{W}_k$, with $k=\lfloor\frac{n-1}{2}\rfloor$, 
 then it uniquely solves \eqref{eq:decoding} in $\mathcal{W}_k$.}
\end{theorem}

\emph{Proof.} 
Since $G$ is bipartite, by Lemma \ref{lem:R_nonbiptt}, any submatrix of $R$  with $n-1$ columns has full column rank.  Hence, by Lemma \ref{lem:compsensing_unique}, if there exists a solution $w\in \mathcal{W}_k$ of \eqref{eq:Rw}, 
then it is unique in $\mathcal{W}_k$. 
The proof ends by noticing that 
if $w$ is a unique solution of \eqref{eq:Rw} in $\mathcal{W}_k$ then it is the unique solution of \eqref{eq:decoding} in $\mathcal{W}_k$ (see Lemma \ref{lem:ifpart}).
 \qedp


{To ensure uniqueness of the solution in \eqref{eq:decoding},   approximately half of the sensors are required  to be bias free by Theorem \ref{thm:suff-condt_biptt}. Next, we introduce rather mild restrictions on the admissible set of biases $\mathcal{W}$ in order to obtain more relaxed conditions on the number of bias free sensors.}
 
\begin{definition}\label{def:het.biases}
\begin{enumerate}[(i)]
	\item The set {$\mathcal{W}_{k}^h$}, with $2\le k\le n$, of heterogeneous $k$-sparse bias vectors is the set of all  vectors $w\in \calW_k$ such that their {nonzero} entries are different from each other, namely $w_i\neq w_j$ for any $i,j\in V$ with $w_i\ne 0$ and $w_j\neq 0$.
	\item The set $\mathcal{W}_{k}^a$, with $2\le k\le n$,  of {\em absolutely} heterogeneous $k$-sparse bias vectors is the set of all vectors {$w\in \calW_k$} such that their {nonzero} entries in {\em absolute value} are different from each other, namely $|w_i|\neq |w_j|$ for any $i,j\in V$ with $w_i\ne 0$ and $w_j\neq 0$.
	\end{enumerate}
\end{definition}

{Note that we have
$
\calW_k^a \subset \calW_k^h \subset \calW_k
$, for each $k=2, 3, \ldots, n$.}
\begin{theorem}\label{thm:heterog-biases}
Consider a bipartite graph $G$,
	\begin{enumerate}[(i)]
		\item If there exists $w$ that solves \eqref{eq:decoding} in $\mathcal{W}^h_{n-3}$, then it uniquely solves \eqref{eq:decoding} in $\mathcal{W}_{n-3}$.
		\item If there exists $w$ that solves \eqref{eq:decoding} in $\mathcal{W}_{n-2}^a$, then it uniquely solves \eqref{eq:decoding} in $\mathcal{W}_{n-2}$.
	\end{enumerate}	

\end{theorem}
\emph{Proof.} 
{Noting Lemma \ref{lem:ifpart}, we work with equation \eqref{eq:Rw} to prove uniqueness of the solution. }

$(i)$ 
We prove this part by contradiction.  Suppose there exists another solution $w'\neq w$ of \eqref{eq:Rw}, satisfying $w'\in \mathcal{W}_{n-3}$. Then 
\begin{eqnarray}\label{eqn:null-space-R}
R(w-w')=0.
\end{eqnarray}
By \eqref{eq:KerR} this implies that $w=w'+ fa$, where $f$ is given by \eqref{eq:f} and $a\in \R$.

%
Let $\mathcal{S}_{w}$ and $\mathcal{S}_{w'}$ be the support of $w$ and $w'$. 
If $V\setminus (\mathcal{S}_{w}\cup\mathcal{S}_{w'})$ is nonempty, i.e,
there exists at least one index $i\in V$ such that $w_i-w'_i=0$, then $a=0$. This implies $w=w'$ and leads to a contradiction. If $\mathcal{S}_{w}\cup\mathcal{S}_{w'}=V$, we have that {$\mathcal{S}_{w}\setminus\mathcal{S}_{w'}= (\mathcal{S}_w \cup \mathcal{S}_{w'})\setminus \mathcal{S}_{w'}=V\setminus \mathcal{S}_{w'}$} should have at least $3$ elements
since\footnote{\label{footnote-set-ex} Note the following two identities: $|\mathcal{S}_{w}\cup \mathcal{S}_{w'}|=|\mathcal{S}_{w}|+|\mathcal{S}_{w'}|-
|\mathcal{S}_{w}\cap \mathcal{S}_{w'}|$ and $|\mathcal{S}_{w'}|= |\mathcal{S}_{w}\cap \mathcal{S}_{w'}|+|\mathcal{S}_{w'}\setminus \mathcal{S}_{w}|$. Replacing the right-hand side of the second identity into the first one, we obtain 
$|\mathcal{S}_{w}\cup \mathcal{S}_{w'}|=|\mathcal{S}_{w}|+|\mathcal{S}_{w'}\setminus \mathcal{S}_{w}|$, or 
$|\mathcal{S}_{w'}\setminus \mathcal{S}_{w}|= |\mathcal{S}_{w}\cup \mathcal{S}_{w'}|- |\mathcal{S}_{w}|$. Since $|\mathcal{S}_{w}\cup \mathcal{S}_{w'}|=n$ and $|\mathcal{S}_{w}|\le n-3$, we obtain $|\mathcal{S}_{w'}\setminus \mathcal{S}_{w}|\ge 3$, as claimed.
}
 $\|w'\|_0\le n-3$.
 However, this would imply that there exist at least three distinct indices $i,j,k\in \mathcal S_{w}\setminus\mathcal{S}_{w'}$, such that each one of 
$w_i$, $w_j$, $w_k$ is either equal to $a$ or $-a$, {with $a\neq 0$}. Hence, at least two elements in  the set $\{w_i, w_j, w_k\}$ must be the same, which contradicts the {heterogeneity assumption $w\in \mathcal{W}_{n-3}^h$}. This completes the proof of uniqueness for part (i).

$(ii)$ Suppose by contradiction that there exists another solution $w'\neq w$ of \eqref{eq:Rw}, satisfying $w'\in \mathcal{W}_{n-2}$.
Analogous to the proof of $a)$, if $V\setminus (\mathcal{S}_{w}\cup\mathcal{S}_{w'})$ is nonempty, then $w=w'$, while if $\mathcal{S}_{w}\cup\mathcal{S}_{w'}=V$, the set $\mathcal S_{w}\setminus\mathcal{S}_{w'}$ has at least $2$ elements since $\|w\|_0\le n-2$. This would imply that there exist at least two distinct indices $i,j\in \mathcal S_{w}\setminus\mathcal{S}_{w'}$, such that each one of $w_i$ and $w_j$ is equal to either $a$ or $-a$, {with $a\ne 0$.} This results in $|w_i|=|w_j|$, thus contradicting the {absolute heterogeneity assumption $w\in \mathcal{W}_{n-2}^a$. This completes the proof.} 
of \eqref{eq:decoding}.
\qedp

\medskip

Thus, focusing the attention on the class of heterogeneous biases in the sense of Definition \ref{def:het.biases} considerably increases the number of allowable biased sensors. 


\subsection{Distributed bias computation with coordinator}\label{subsec:ctcoordinator}


In this subsection we focus on algorithms for computing the actual vector of biases $w$.  We propose the use of a coordinator that delegates the computation of the biases to the nodes while organising the execution of their commands. Compared to a centralized solution, the distributed computation with a coordinator eases the analysis and does not require to know the network topology. 

We consider the case when $w\in \mathcal{W}_{n-2}^a$ and use the result established in Theorem \ref{thm:heterog-biases} (ii).
When $w\in \mathcal{W}_{n-2}^a$, there exist at least two (bias free) nodes $i,j\in V$, $i\ne j$, satisfying $w_i=w_j=0$. 
{The essence of the algorithm here is to find such a bias free pair.   To this end, some additional notation is needed.} For a pair of nodes $i,j\in V$ with $i\ne j$, let $\mathcal{P}_{ij}$ be {a} path connecting them,  namely $\mathcal{P}_{ij}=\{k_0, k_1, \ldots, k_{d_{ij}}\}$, with $k_0=i$, $k_{d_{ij}}=j$, and $d_{ij}$  the length of the path.   
{Moreover, we collect the 
{measurements}
that are indexed by $\mathcal{P}_{ij}$ as}
\be\label{Z.ij}
Z_{ij}:=
\begin{bmatrix}
	z_{k_0\, k_1}+z_{k_1\, k_0}\\
	z_{k_1\, k_2}+z_{k_2\, k_1}\\
	\vdots\\
	z_{k_{d_{ij}-1}\, k_{d_{ij}}}+z_{k_{d_{ij}}\, k_{d_{ij}-1}}
\end{bmatrix}.
\ee
Finally, we let
\be\label{vec.e}
 e_{d_{ij}}= \begin{bmatrix} 
	(-1)^{d_{ij}-1} & (-1)^{d_{ij}-2} & \ldots & (-1)^1 & (-1)^0 
\end{bmatrix}^\top.
\ee
We then have the following result:
\begin{proposition}\label{prop:Iij}
	Consider a bipartite graph $G$,
{let $w$ be the vector of biases and assume that $w\in \mathcal{W}_{n-2}^a$.}
For a given pair of nodes $i,j\in V$, with $i\ne j$,  and a path $\mathcal{P}_{ij}$ connecting them, we have:

 \begin{enumerate}[(i)]
		\item $I_{ij}:=e_{d_{ij}}^\top Z_{ij}=0$ if and only if $w_i=w_j=0$, i.e., the pair $i,j\in V$ is bias-free.
		\item If $w_i=0$, then $I_{ij}=w_j$.  
		\item  $I_{ik_{\ell}}=-I_{ik_{\ell-1}}+(z_{k_{\ell-1}k_{\ell}}+z_{k_{\ell}k_{\ell-1}})$ for $\ell\in \{2,...,d_{ij}\}$, where $I_{ik_{\ell}}, I_{ik_{\ell-1}}$ are defined similarly to $I_{ij}$.
	\end{enumerate}
\end{proposition}

\emph{Proof.}  
(i) 
{By \eqref{eq:biased_meas},} 
the vector $Z_{ij}$ equals 
\be\label{Z.ij.explicit}
Z_{ij}=
\begin{bmatrix}
	w_{k_0}+w_{k_1}\\
	w_{k_1}+w_{k_2}\\
	\vdots\\
	w_{k_{d_{ij}-1}} +w_{k_{d_{ij}}}
\end{bmatrix}
\ee
from which 
\be\label{boh}
\ba{rl}
I_{ij}=& e_{d_{ij}}^\top Z_{ij}= \sum_{\ell=1}^{d_{ij}} (-1)^{d_{ij}-\ell} (w_{k_{\ell-1}}+w_{k_{\ell}})\\
=& (-1)^{d_{ij}-1} w_{k_0} +w_{k_{d_{ij}}}\\
=& (-1)^{d_{ij}-1}w_i +w_j.
\ea
\ee
Noting  $w\in \mathcal{W}^a_{n-2}$, we find that  $e^\top Z_{ij}=0$ if and only if $w_i=w_j=0$, as claimed. 

(ii)  By \eqref{boh}, we immediately obtain that {$I_{ij}= w_j$ if $w_i=0$.}

(iii) The conclusion is straightforward to obtain by the definition of $I_{ij}$ and \eqref{vec.e}.
\qedp

From Proposition \ref{prop:Iij} (i), no matter along which path the quantity $I_{ij}$ is computed, the identity $I_{ij}= 0$ holds if and only if the pair $i,j\in V$ is bias-free. 
Hence, $I_{ij}$ is an indicator of whether or not a pair of nodes are bias free. 
In addition, by Proposition \ref{prop:Iij} (iii), if node $k\in \mathcal{N}_j$ knows $I_{ij}$, then it can compute $I_{ik}$. In turn, by Proposition \ref{prop:Iij} (ii), if $w_i=0$, then the variable $I_{ik}$ equals the bias $w_k$. Based on Proposition \ref{prop:Iij}, searching the bias free nodes and solving the bias can be concurrently carried out by the nodes in a distributed fashion coordinated by a coordinator. The idea is to let the coordinator make $n-1$ selections of a candidate bias-free node $i$ and let the other nodes $j$ compute the variables $I_{ij}$ with respect to the selected node. {As soon as a zero $I_{ij}$ is observed at a node $j$, then that node informs the coordinator to terminate the search. At this stage,  every node has computed  the value of its bias via the indicator variable, namely $I_{ij}=w_j$.}

\begin{algorithm}[t]\caption{Coordinator }\label{algrm:A}
	\textbf{Data:} Set of nodes $V$ and counter $T$\; 
	\textbf{Initialize:} $T:=0$\; 
	\For {$i=1: n-1$}{
		Inform all the nodes in $V$ to start the {\bf Node pair test stage} in Algorithm 2\;
		Inform node $i$ that it is selected and nodes $j\in V\setminus{i}$ that they need to calculate and send back the variable $I_{ij}$ to the coordinator\; 
		$T=T+1$\;
		Once {\bf Node pair test stage} is completed by all the nodes, receive  $I_{ij}$ and $t_j$ from all $j\in V\setminus{i}$\;
		Compute $T=T+\max_{j\in V \setminus\{i\}} \{t_j\}$\;
		\If{there exists one $I_{ij}=0$} 
		{
		{Stop the {\bf for} iteration}\;   	}	   				
	}	
	{ 
	Inform all the nodes to start the \textbf{Bias computing stage}}\;			
\end{algorithm}
\begin{algorithm}[t]\caption{Node $j$}\label{algrm:B}
	\textbf{Data:} Set of neighbors $\mathcal{N}_{j}$, measurement data $\{ z_{jk}+z_{kj}\}_{k\in\mathcal{N}_j}$ and counter $t_j$\;
	\uIf{ informed to start the \bf{Node pair test stage}}{
				{\tcc{\bf Node pair test stage}}				
				\uIf{node $j$ is selected in iteration $i$, i.e. $j=i$
				}{
				Set the auxiliary variable $I_{jj}=0$ and $t_j=1$\;
				Send $(I_{jj}, t_j)$ to all $k\in \mathcal{N}_j$\;
				Stop accepting data from the neighbors\;
			}
			\Else {Once $(I_{ik}, t_k)$, for some $k\in \mathcal{N}_j$, are received, pick any one of $(I_{ik}, t_k)$ and compute $I_{ij}:=-I_{ik}+(z_{jk}+z_{kj})$, $t_j=t_k+1$\;
				Send $(I_{ij}, t_j)$ to all $k\in \mathcal{N}_j$ and the coordinator\; 		
				Stop accepting data from the neighbors\;
			}}
			\If{  informed to start the \bf{Bias computing stage}}
				{\tcc{\bf Bias computing stage} 
				$w_j=I_{ij}$\;}
\end{algorithm}

The commands executed by the coordinator are summarized in Algorithm \ref{algrm:A}, whereas the commands executed by the nodes are listed in Algorithm \ref{algrm:B}. Algorithm \ref{algrm:B} comprises two stages, \emph{the node pair test stage}, in which the coordinator and the nodes cooperate to check whether or not a given pair of nodes is bias free, and \emph{the bias computing stage} during which the biases are explicitly computed. In Algorithm \ref{algrm:B}, we assume that each node has access to the data $\{z_{ij}+z_{ji}\}_{j\in \mathcal{N}_i}$, which can be achieved by letting all the nodes collect the measurements from their neighbors, {before running Algorithms \ref{algrm:A} and \ref{algrm:B}.} 

To measure  
the number of executed instructions required by the algorithms to terminate the computation,
we introduce counters that store integer values.
{
In Algorithm \ref{algrm:A}, the sequence of actions by the coordinator consisting of informing node $i$ that it has been selected, and asking nodes $j\in V\setminus{i}$ to calculate and send back the variable $I_{ij}$ is considered as one instruction, which increases the counter $T$ by $1$ unit.  The single action of informing all the nodes to start the \textit{Bias computing stage}, is regarded as another instruction, and again results in an increase of $T$ by $1$ unit. }
In Algorithm \ref{algrm:B}, at each iteration $i$, the variable $t_j$, $j\in V$, stores the number of instructions executed {from the moment  that node $i$ is selected by the coordinator till when $j$ computes $I_{ij}$.} The counters $t_j$, $j\in V$, are communicated to the coordinator and used to update the counter $T$, which therefore contains  the {total} number of instructions executed before the bias free node pair is found. {Note that the counters are only introduced to store the number of instructions needed for the computation of the solution, as formalized in Theorem \ref{thm:finitestep}, but do not play any role in the computation of the solution itself.}

The following result summarizes the properties of the algorithms:
\begin{theorem}\label{thm:finitestep}
	Consider a bipartite graph $G$, with its diameter given by $D_G$,
{let $w$ be the vector of biases and assume that $w\in \mathcal{W}_{n-2}^a$.}
	If the coordinator uses Algorithm \ref{algrm:A} and the nodes Algorithm \ref{algrm:B}, then {a bias free node can be identified in $T$ instructions} and the vector of biases $w$ can be {reconstructed} in $T+2$ instructions with $T\le (n-1)(D_{G}+2)$.
\end{theorem}

\medskip

\emph{Proof.} At iteration $i$, with $i=1, 2, \ldots, n-1$,   the coordinator selects node $i$ and  informs all the nodes to start the node pair test stage (see Algorithm \ref{algrm:A}). We first focus on the node pair test stage.


According to Algorithm \ref{algrm:A}, if 
node $i\in V\setminus\{n\}$ is selected, the coordinator informs all the nodes $k\in V$,  
{and $T$ is increased by $1$}. According to Algorithm \ref{algrm:B}, when node $i$ receives the message from the coordinator that it has been selected, it sets $I_{ii}=0$ and $t_i=1$, and sends them to all the neighbors $j\in \mathcal{N}_i$. 
{The instructions executed from the instant when node $i$ has been informed that it has been selected to the instant when nodes $i$ computes $I_{ii}$ are regarded as one and it is set $t_i=1$.} 

When the node $j\in \mathcal{N}_i$ receives $(I_{ii},t_i)=(0,1)$, it computes $I_{ij}=-I_{ii}+(z_{ij}+z_{ji})=I_{ij}$ and $t_j=t_i+1=2$, then sends $(I_{ij}, t_j)$ to the coordinator and its neighbors. Hence $t_j=2$ actions are executed from the instant when node $i$ is informed to have been selected to the instant when node $j$ computes $I_{ij}$. 
{ Let $D_{i}^{{\rm max}}$ be the maximum of the distances of node $i$ to all}  other nodes in $V$.  
Consequently, each node $j_{\ell}$, which is at a distance $\ell\in \{2,3,...,D_i^{{\rm max}}\}$ from  node $i$, receives $(I_{ij_{\ell-1}}, t_{j_{\ell-1}})$,
with $t_{j_{\ell-1}}=\ell$, from some neighbor $j_{\ell-1}$, which is at a distance $\ell-1$ from node $i$. The node $j_{\ell}$ computes $t_{j_\ell}=t_{j_{\ell-1}}+1=\ell +1$ and, in view of Proposition \ref{prop:Iij} (iii), we have
\begin{eqnarray}
I_{ij_{\ell}}&=& -I_{ij_{\ell-1}}+( z_{j_{\ell-1}j_{\ell}}+z_{j_{\ell}j_{\ell-1}}).
\end{eqnarray}
All the nodes $j_{\ell}$ then send $(I_{ij_{\ell}}, t_{j_{\ell}})$, with $t_{j_{\ell}}=\ell+1$,  to their neighbors and the coordinator. Hence $t_{\ell}=\ell+1$ instructions are executed  from the instant when node $i$ is informed it has been  selected to the instant when node $j_{\ell}$ computes $I_{ij_{\ell}}$. By this analysis,  after node $i$ has been informed at iteration $i=1,2,...,n-1$, $t_{D_{i}^{{\rm max}}}=\max_{j\in V\setminus\{i\}}\{t_{j}\}=D_{i}^{{\rm max}}+1$ instructions are executed before the coordinator receives $I_{ij}$ from all $j\in V\setminus\{i\}$. Hence at each iteration $i=1,2,...,n-1$, $T$ is increased of at most 
to check where the extra $+1$ comes from

Since $I_{ij}$ and $I_{ji}$ can be used interchangeably, the coordinator obtains all $I_{ij}$ for $i,j\in V$, $i\ne j$, in at most $n-1$ iterations. By the assumption $w\in {\mathcal{W}^a_{n-2}}$ and Proposition \ref{prop:Iij}, there always exists an iteration $i=1,2,...,n-1$ and a node $j\in V\setminus\{i\}$ such that $I_{ij}=0$. Hence the bias free node pair should be found in $T\le (n-1)(D_G+2)$ steps.

We then consider the bias computing stage. This occurs if the coordinator received $I_{ij}=0$ at iteration $i$ for some $j\in V\setminus\{i\}$. Then each node $k\in V$ enters this stage and it concludes that the computed quantity $I_{ik}$ is the bias $w_k$. As a matter of fact, since $I_{ij}=0$, then $w_i=0$, by Proposition \ref{prop:Iij} (i),  and this actually implies that $I_{ik}=w_k$ if $k\ne i$, by Proposition \ref{prop:Iij} (ii). For $k=i$, we note that $I_{ii}$ was set equal to zero in  the node pair test stage, and therefore $I_{ii}= w_i=0$. 

To complete the computation of the number of executed instructions, we note that by Algorithm \ref{algrm:B} one more instruction is needed to let the coordinator inform all the nodes that $i$ is bias free and another instruction to let the nodes compute the biases. 
\qedp

A few remarks are in order:
\begin{enumerate}[-]
	\item 
	{In case $w\in \mathcal{W}_{n}^a\setminus \mathcal{W}_{n-2}^a$, so that  the assumption $w\in \calW_{n-2}^a$ in Theorem \ref{thm:finitestep} is not satisfied, then $I_{ij}= 0$ will not be observed at any node, and the coordinator infers that  
	there is no pair of bias-free nodes.} 
\item 	
In Algorithm 1, the coordinator is only responsible for coordinating the nodes{, namely initializing each iteration, whereas all computations are performed at the nodes in a distributed fashion.}  Moreover, note that the coordinator does not need to know the topology of the network, {apart from the node set $V$.}
\item 	Another method to compute the  vector of biases when $w\in \mathcal{W}^a_{n-2}$ is to combinatorially search the pair of nodes that is bias-free, as in \cite{Fawzi2014SecureEA,lee2015secure}. Specifically, for each pair of indices $i,j\in V$, with $i\ne j$, one could look for a solution of the modified equation $Rw^{(i,j)}=\tilde z$, where $w^{(i,j)}$ 
is a vector whose entries $i$ and $j$ are set to zero. If a solution to this modified equation exists,  then by construction it satisfies the sparsity condition $\|{w}^{(i,j)}\|_0\le n-2$, and by Theorem \ref{thm:heterog-biases} (ii), {it will be equal to the vector of actual biases.}  Hence, the determination of the vector of biases $w$ satisfying  \eqref{eq:decoding} is reduced to considering the $n(n-1)/2$ systems of equations and check if each of these equations admit a solution. Note however that such an approach would require that the unit carrying out the combinatorial search has access to the network topology and possesses enough computational power. 

\end{enumerate}


\subsection{Distributed bias estimation without coordinator}

In the previous section 
we assumed the existence of  a coordinator that  supervises the nodes  checking the conditions of Proposition \ref{prop:Iij}. 
In this section, we seek  a method that estimates the biases in a distributed manner without resorting to a coordinator. We show that this is achievable provided that we restrict the class of admissible biases.
To this end, by Subsection \ref{subsec:decoding} and equation \eqref{eq:Rw}, we consider the following $\ell_1$-norm minimization problem
\begin{eqnarray}\label{eq:biases_ell_1min}
\min_{\overline w\in \mathbb R^n} && \|\overline w\|_1\\
\mathrm{s.t.} && R \,\overline w=\tilde z,\nonumber
\end{eqnarray}
where $\tilde z$ is the vector of known values appearing in \eqref{eq:Rw}.
As mentioned in Section \ref{subsec:decoding}, solving the $\ell_1$-norm minimization problem may yield a solution 
{that is different from the vector of actual biases $w$. The sparsity condition under which the solution of \eqref{eq:biases_ell_1min} coincides with $w$ is provided in the following theorem.}

\begin{theorem}\label{thm:bias_ell_1min}
{For a  bipartite graph $G$, the vector of  biases $w$} is the unique solution of the $\ell_1$-norm  minimization problem \eqref{eq:biases_ell_1min} if 
the number of biased sensors is not greater than $\lfloor\frac{n-1}{2}\rfloor$, i.e., $w\in \mathcal{W}_{\lfloor\frac{n-1}{2}\rfloor}$.
\end{theorem} 
\emph{Proof.} 
{Since the graph is bipartite, then \eqref{eq:KerR} holds. Hence,} 
{inequality \eqref{eq:nup} in this case is given by} 
\begin{eqnarray}
\sum_{i\in S,|S|=s} |v_i| &<& \sum_{j\in S_c} |v_j|\nonumber\\
\Longleftrightarrow\qquad  s|a| &<& (n-s)|a|, {\quad a\ne 0}\label{nsp}
\end{eqnarray}
which is satisfied {if and only if} $s<\frac{n}{2}$. {Hence, the matrix $R$ satisfies the null space property of order $s$, with $s=  \lfloor\frac{n-1}{2}\rfloor$.}

Therefore, by \eqref{eq:Rw}, Theorem \ref{thm:nup_recovery} {and the discussion following it, if the  vector of biases $w$ in \eqref{eq:Rw} satisfies $w\in \mathcal{W}_{\lfloor\frac{n-1}{2}\rfloor}$, then there exists a unique solution of the optimization problem \eqref{eq:biases_ell_1min}, with $\tilde z = Rw$, and it is equal to $w$. 
%
\qedp

\smallskip

This theorem shows that for bipartite graphs, the $\ell_1$-norm minimization does not decrease the maximum number of allowed biased sensors {obtained} in {Theorem \ref{thm:suff-condt_biptt}.} 
{On the other hand, in the case where the vector of biases $w$ belongs to the {set of heterogeneous biases $\mathcal{W}_{n-3}^h$ or $\mathcal{W}_{n-2}^a$ 
considered in Theorem \ref{thm:heterog-biases},}
examples can be found where the solution of the $\ell_1$-norm minimization problem does not give the correct bias estimation. Hence, below, we only discuss the solution of \eqref{eq:biases_ell_1min} for the case of bipartite graphs with a number of biased sensors as characterized in Theorem \ref{thm:bias_ell_1min}.
}
%

The $\ell_1$-norm optimization problem \eqref{eq:biases_ell_1min} can be solved directly in a distributed manner by the methods in \cite{Jinqiu_DistL1min,Mota2012DistributedBP}. In this paper, 
we reformulate it as a linear programming problem as \cite{ChenSS_BasisPursuit}
\begin{eqnarray}\label{eq:LP}
\min_{\eta\in \mathbb{R}^{2n} } && \mathbb 1_{2n}^\top \eta\\
\mathrm{s.t.} && H\eta=\tilde z,\ \eta\ge 0 \nonumber
\end{eqnarray}
where $\eta$ is the 
{decision}
variable and 
\begin{eqnarray}\label{eq:cH}
H=\begin{bmatrix}R & -R \end{bmatrix}.
\end{eqnarray}
{Under the sparsity condition in Theorem \ref{thm:bias_ell_1min},} if $\eta^*$ is the solution of \eqref{eq:LP}, the vector of biases can be computed as
\begin{eqnarray}\label{eq:weta}
{w=\begin{bmatrix}I_n & - I_n \end{bmatrix}}
 \eta^*
\end{eqnarray}

{The linear programming problem above can be solved by various distributed methods available in the literature, 
see e.g. \cite{Feijer2010StabilityOP,Brger2012ADS,RichertD_Dist_LP}. In particular, using the result of \cite{RichertD_Dist_LP}, the bias estimation algorithm takes the form}
\begin{eqnarray}\label{eq:dist_LP}
\hat w &=&
{\begin{bmatrix} I_n & - I_n\end{bmatrix}}
\eta\nonumber \\[1mm]
\dot \eta_i&=&\left\{\begin{array}{cc}
f_i(\eta,\lambda),& \mathrm{if}\ \eta_i>0\\[1mm]
\max\{0,f_i(\eta,\lambda)\}& \mathrm{if}\ \eta_i=0
\end{array} 
\right., \  i\in V \label{eq:distLP_eta}\nonumber\\
\dot \lambda &=& H\eta-\tilde z \label{eq:distLP_lambda}\nonumber\\[-3mm]
\end{eqnarray}
with
\begin{eqnarray}\label{eq:distLP_f}
f(\eta,\lambda)=-\mathbb{1}_{2n}-H^\top(\lambda+H\eta-\tilde z),
\end{eqnarray}
and where $\lambda\in \R^m$ is the dual variable and the initial condition satisfies $\eta_i(0)\ge 0$ for all $i\in V$.
%
%

For this algorithm, we have the following result:
\begin{proposition}\label{thm:bip_asympstable}
The estimate $\hat w$ generated by the algorithm \eqref{eq:dist_LP}, \eqref{eq:distLP_f} converges asymptotically to the vector of biases $w$ if $G$ is bipartite and $w\in \mathcal{W}_{\lfloor\frac{n-1}{2}\rfloor}$. 
\end{proposition}
\emph{Proof}. This result follows directly from \cite[Proposition IV.4]{RichertD_Dist_LP}  noting that the linear program \eqref{eq:LP} has a unique solution. \qedp

\begin{remark}
\rm{ Similarly to Subsection \ref{subsec:bias.rejection}, one could use the estimate $\hat w$ generated by the algorithm \eqref{eq:dist_LP}, \eqref{eq:distLP_f} in the compensator \eqref{eq:compensator} to reject the effect of the biases and achieve consensus. In fact, the consensus dynamics \eqref{eq:dynbiaserr} driven by the estimation error $e$ continues to be valid and an analysis similar to the one in Proposition \ref{thm:consensus} can be carried out. In the case of bipartite graphs, however, we cannot provide the estimate of the new consensus value, due to the lack of the exponential convergence of the estimation error.}\qedp
\end{remark}

\smallskip

	For the problem at hand,  the algorithm \eqref{eq:dist_LP} has some advantages when compared with possible alternatives, such as the one provided by the recent paper \cite{Jinqiu_DistL1min}, where a new distributed algorithm for solving the $\ell_1$-norm minimization problem with linear equality constraints is proposed. 
	{However, in this method each node needs to reconstruct all the elements of the solution of the $\ell_1$-norm minimization problem,  which implies that each node stores and communicates a vector with the same dimension as the  (unknown) solution. Moreover, an implicit}
	requirement for the method in \cite{Jinqiu_DistL1min} is that each agent must know the number of columns of the coefficient matrix{, which translates to knowing the network size in our setting.}
		In the method given by \eqref{eq:dist_LP}, on the other hand, each  node reconstructs  only one element of $w$ by communicating suitable variables with its neighbor. The latter is done without relying on any global information including the size of the network.}
	
\begin{remark}
\rm{Resorting to different formulations of the $\ell_1$-norm minimization problem, one can obtain variations of the algorithm \eqref{eq:dist_LP} with different features. 
For instance, 
	\eqref{eq:biases_ell_1min} can be reformulated as
	\begin{eqnarray}
	\min_{\overline w\in \mathbb{R}^n} && \|\overline w\|_1\\
	\mathrm{s.t.} && R^\top R  \,\overline w=R^\top \tilde z,\nonumber
	\end{eqnarray}
where $R^\top R=A+D$ is the signless Laplacian matrix (see Section \ref{sec:graph.not}). 
	We can transform the above  into a linear {program} analogous to \eqref{eq:LP}.  
	Then, one can write a distributed algorithm similar to \eqref{eq:dist_LP} 
	for which the variable $\lambda$ is now defined on the nodes, and thus has  $n$ elements.
However, $H$ in \eqref{eq:cH} becomes $[R^\top\;  R\; -R^\top\; R]$. The term $H^\top H\eta$ in \eqref{eq:distLP_f} requires each node $i$ to collect not only $\eta_j$, for $j\in \mathcal{N}_i$,  but also $\eta_{k}$, for $k\in \mathcal{N}_j$, which is a two-hop information. On the other hand, in \eqref{eq:dist_LP} each node only needs the decision variables and the dual variables of its neighbors. } \qedp
\end{remark}

\begin{figure}
	\centering
	\includegraphics[scale=0.5]{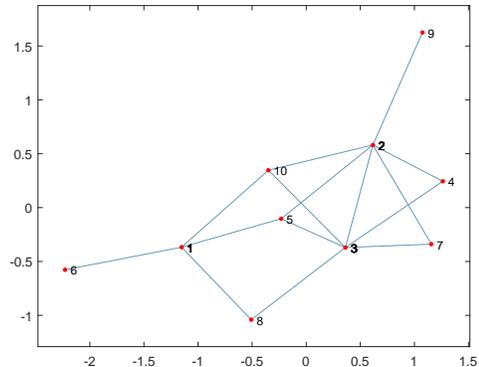}
	\caption{Non-bipartite graph with 10 nodes}
	\label{fig:Non-bip}
\end{figure}
\begin{figure*}
	\centering
	\subfigure[]{\label{fig:x_nbp_ncomp}
		\includegraphics[scale=0.39]{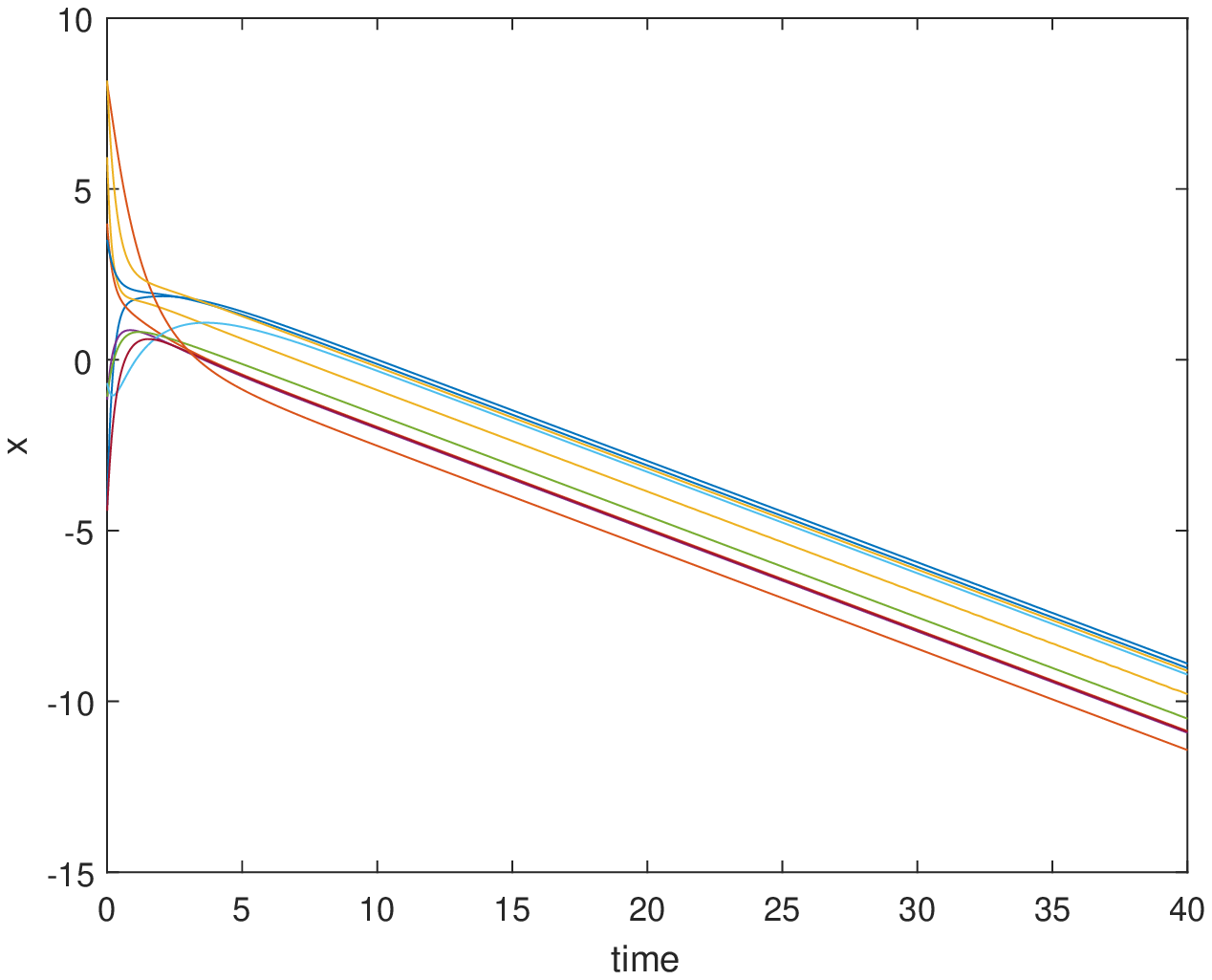}}
	\subfigure[]{\label{fig:e_nbpp_comp}
		\includegraphics[scale=0.39]{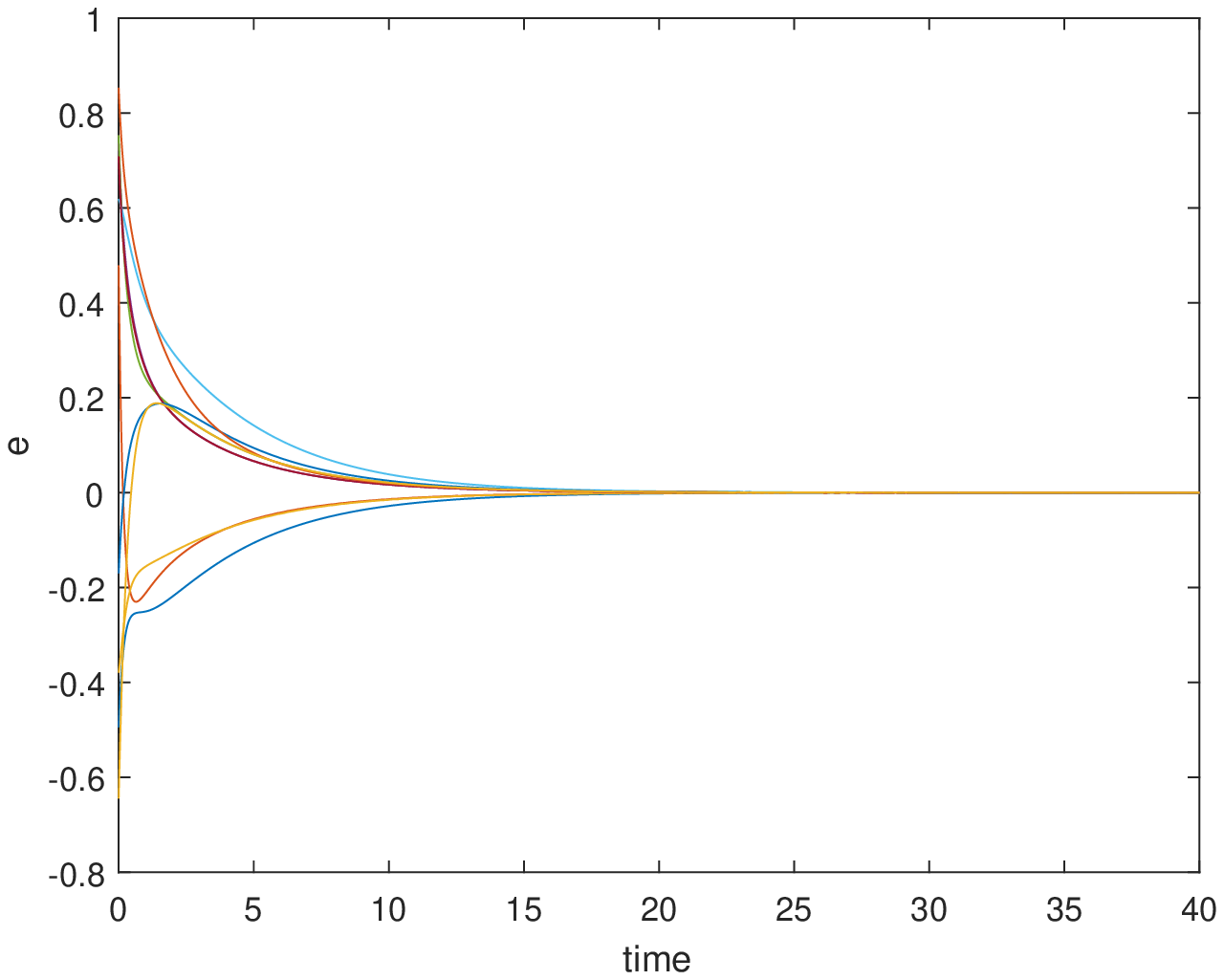}}
	\subfigure[]{\label{fig:x_nbp_comp}
		\includegraphics[scale=0.39]{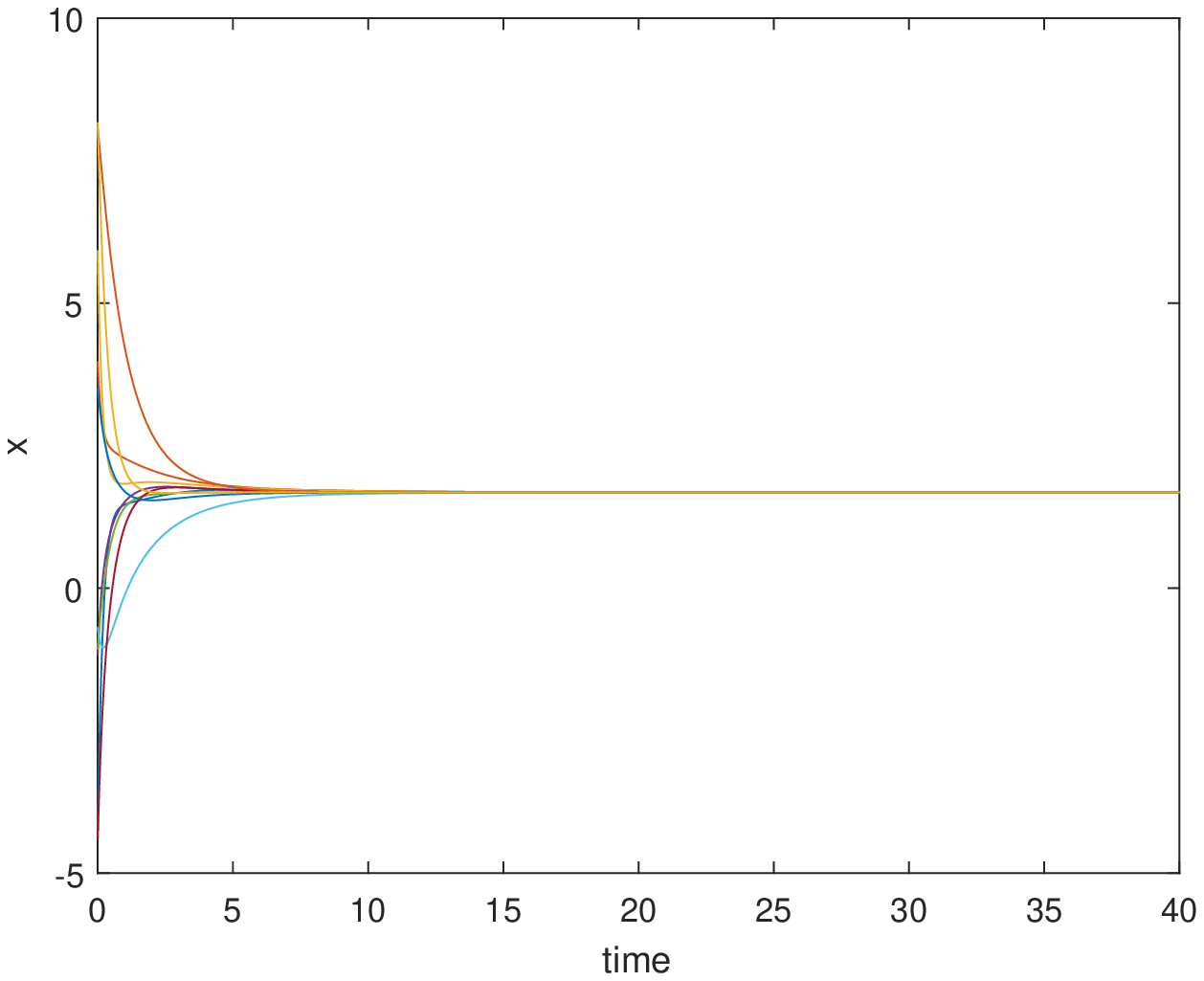}}
	\caption{Bias estimation and consensus evolution for a non-bipartite graph. (a) State evolution of the consensus dynamics \eqref{eq:dyn_bias} without bias compensation; (b) bias estimation error $e$ generated by bias estimator \eqref{eq:biasest}; (c) state evolution of the consensus dynamics \eqref{eq:dyn_bias} with bias compensator \eqref{eq:compensator}.}  	
	\label{fig:nbp}
\end{figure*}

\section{Numerical Simulations} \label{sec:examp}


In this section, we provide numerical simulations 
to illustrate the results for  bias estimation and compensation 
for both non-bipartite graph and bipartite graph.
\subsection{Non-bipartite graphs}
We consider a 
 network with $10$ nodes and each node takes a sensor. The associated graph is non-bipartite and given by Fig. \ref{fig:Non-bip}. The initial state $x_i(0)$ and the bias $w_i$ of each node are generated randomly within the intervals $[-10,10]$ and $[-1,1]$, respectively. A specific example is given as below 
\begin{eqnarray*}
x(0) &=& [-4.280 \ 3.983 \ 5.925 \ -1.168\ -1.076, \\
&& \ -0.687 \ -4.419\ 3.508\ 8.073 \ 8.171]^\top\\
w &=& [0.494\ -0.479\ 0.379\ -0.736\ -0.753\\
&&\ -0.618\ -0.709\ 0.170\ -0.853\ 0.645]^\top
\end{eqnarray*}

We simulate the consensus dynamic \eqref{eq:dyn_bias} with the bias estimator \eqref{eq:biasest} and the bias compensator \eqref{eq:compensator}, where the initial condition for the bias estimate is $\hat w=\mathbb{0}_{10}$. The simulation result is provided in Fig. \ref{fig:nbp}, where Fig. \ref{fig:x_nbp_ncomp} and \ref{fig:x_nbp_comp} show the system state evolution without bias compensation and with bias compensation, respectively, and Fig. \ref{fig:e_nbpp_comp} shows the bias 
estimation error $e$. {As can be seen in Fig. \ref{fig:x_nbp_ncomp}}, if the biases are not compensated, the nodes will not achieve exact consensus and the state of each node $x_i$  drifts away under the influence of the measurement biases. On the contrary, using the bias estimator \eqref{eq:biasest} and the compensator \eqref{eq:compensator}, the bias error $e$ vanishes  and all $x_i$ variables converge to the same  finite value.

\begin{figure}
	\centering
	\advance\leftskip-1cm
	 \advance\rightskip-1cm
	\subfigure[]{\label{fig:e_bp_comp}
		\includegraphics[scale=0.33]{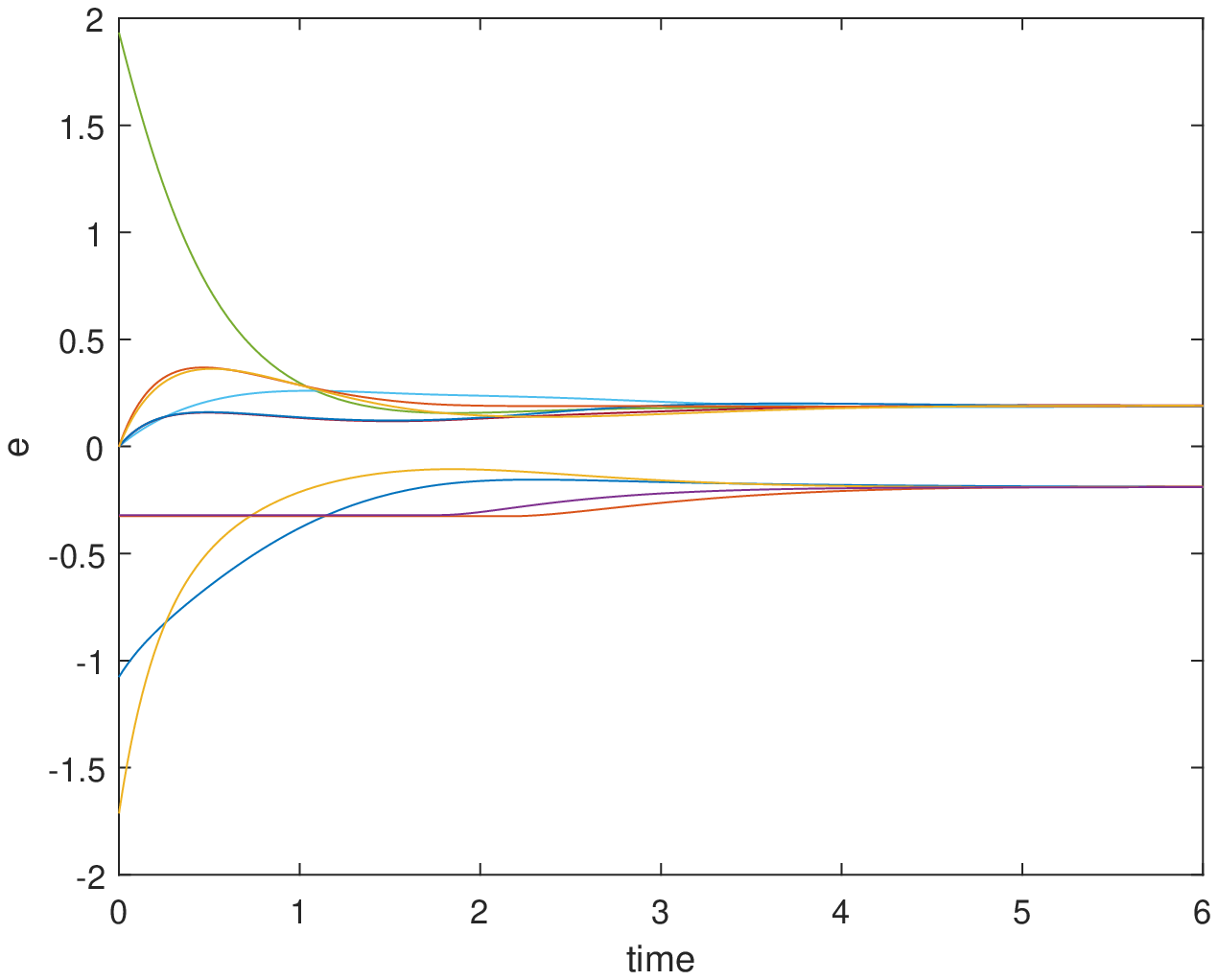}}
	\hspace{-0.3in}
	\subfigure[]{\label{fig:x_bp_comp}
		\includegraphics[scale=0.33]{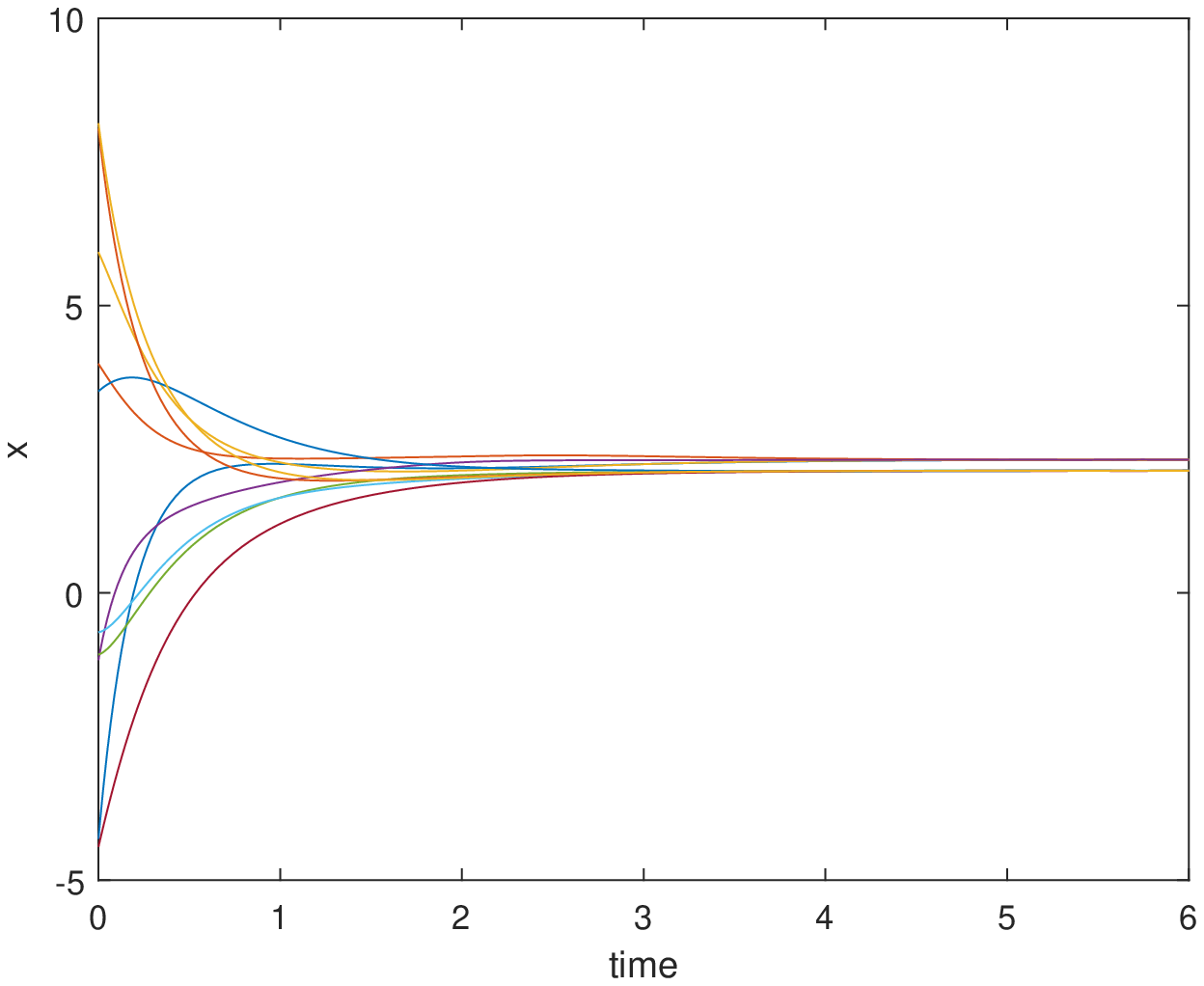}}
	\caption{Bias estimation and consensus evolution for a $10$  node bipartite graph with $5$ biased sensors, hence the condition of Theorem \ref{thm:bias_ell_1min} is violated. The nodes apply the bias estimator \eqref{eq:dist_LP} and the bias compensator \eqref{eq:compensator}.  (a) Bias estimation error; (b) state evolution. }  	
	\label{fig:bp_nonconvg}
\end{figure}

\subsection{Bipartite graphs}
Now, we consider a bipartite graph, which is obtained from the graph in the last subsection removing the edge $\{2,3\}$.
The initial state of the system is the same as the one in the previous subsection.


We first show that if more than $\lfloor\frac{n-1}{2}\rfloor$ sensors of nodes are biased, the $\ell_1$ minimization \eqref{eq:biases_ell_1min} may fail to find the vector of the actual biases $w$ for bipartite graphs. We assume that the sensors of the first five nodes are biased and
\begin{eqnarray}\label{eq:w_exam_bp}
w =[1.076\ 0.326\ 1.713\ 0.320\ -1.932\ 0\ 0\ 0\ 0\ 0]^\top
\end{eqnarray}
We  simulate the consensus dynamics \eqref{eq:dyn_bias} with the bias estimator \eqref{eq:dist_LP} and the bias compensator \eqref{eq:compensator}. The initial conditions for $\eta$ and $\lambda$ are set to zero. The result is given in Fig. \ref{fig:bp_nonconvg}, from which one can see that the entries of the bias estimation error $e$ converge to two values with the same absolute value but opposite signs, thus the biases are not correctly estimated and consensus is not achieved. 

\begin{figure}
	\centering
		\advance\leftskip-1cm
		\advance\rightskip-1cm
	\subfigure[Bias estimation error $e$]{\label{fig:e_bp_lp}
		\includegraphics[scale=0.33]{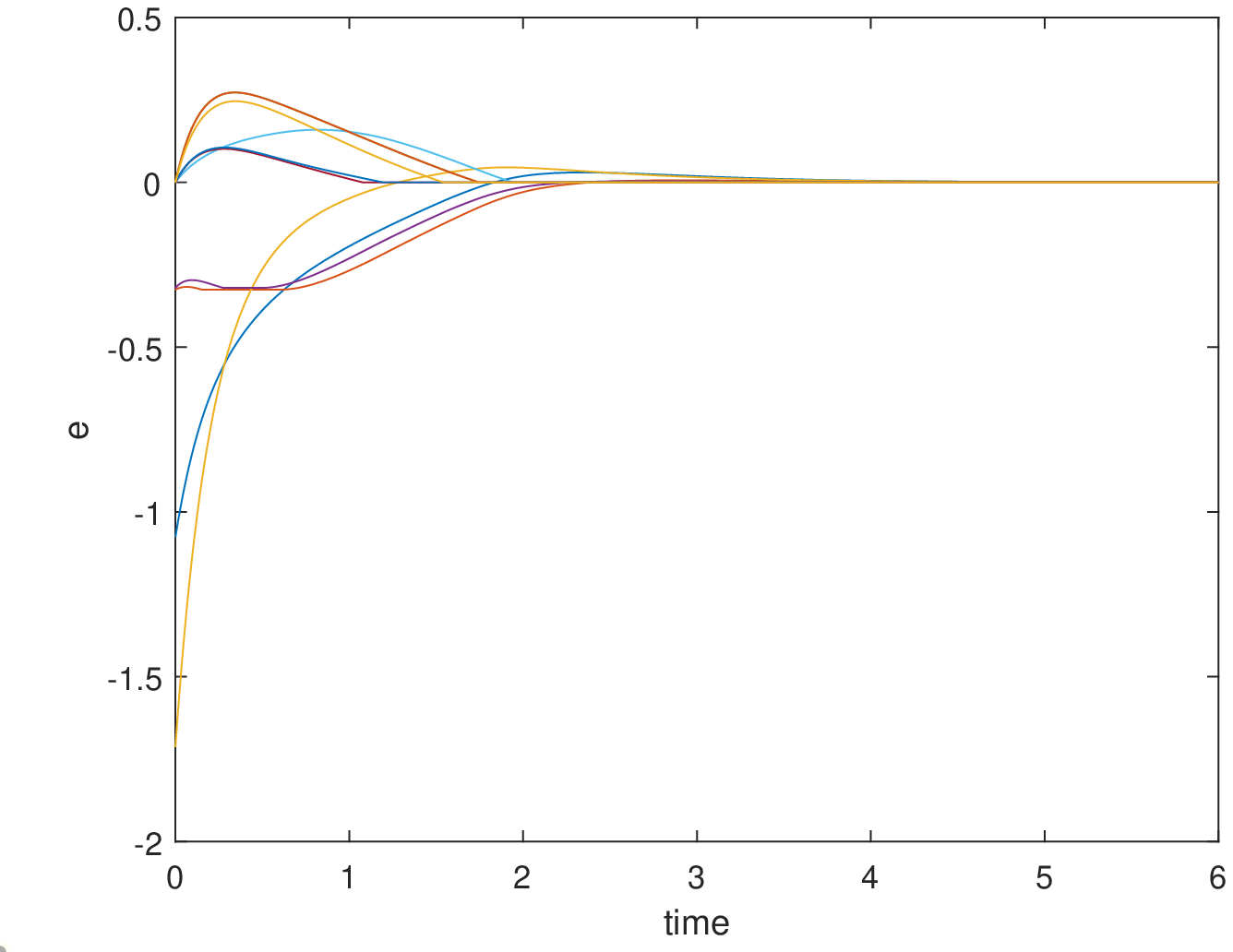}}
	\hspace{-0.3in}
	\subfigure[State evolution with compensation]{\label{fig:x_bp_lp}
		\includegraphics[scale=0.33]{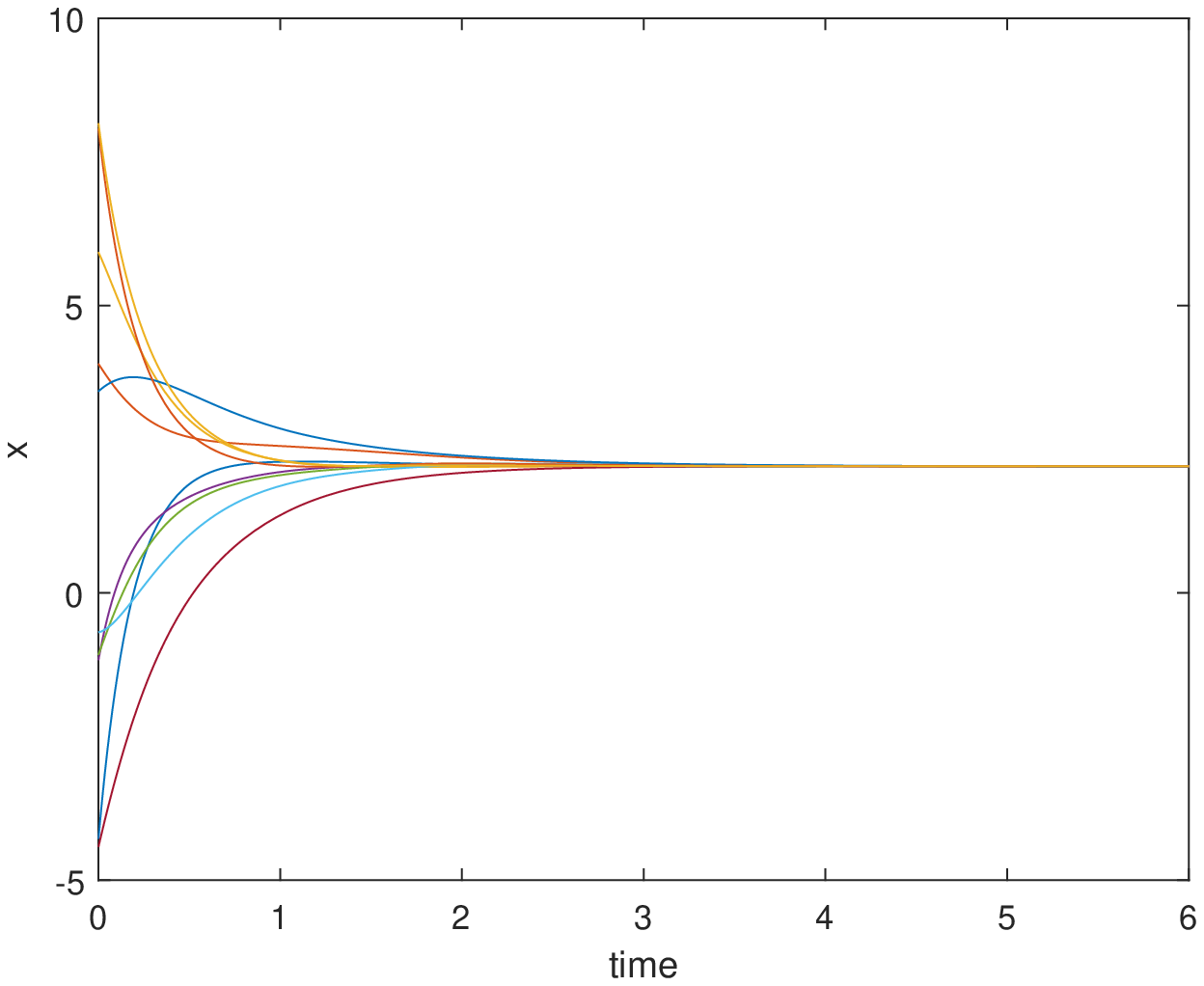}}
	\caption{Bias estimation and consensus for a $10$ node bipartite graph with $4$ biased sensors, hence the condition of Theorem \ref{thm:bias_ell_1min}.   The nodes apply the bias estimator \eqref{eq:dist_LP} and the bias compensator \eqref{eq:compensator}.  (a) Bias estimation error; (b) state evolution.}  	
	\label{fig:bp_convg}
\end{figure}

We then let the sensor of the fifth node also to be unbiased, namely the last six entries of $w$ in \eqref{eq:w_exam_bp} are all zero. The condition of Theorem \ref{thm:bias_ell_1min} is now satisfied. 
The result is depicted in Fig. \ref{fig:bp_convg}, which shows that the bias {estimation} error decays to {zero} and the system achieves consensus. 


\section{Conclusion} \label{sec:conclusion}

In this paper, 
we studied the problem of estimating the biases in sensor networks from relative state measurements, {with an application to the problem of consensus with biased relative state measurement.} Without any sparsity constraint on the biases, we show that the biases can be accurately estimated if and only if the graph is non-bipartite. For bipartite graphs, we show that the biases can be uniquely determined from the measurements if less than half of the sensors is biased. The number of biased sensors can be increased when the biases are heterogeneous, i.e., different from each other, or absolutely heterogeneous, i.e., with absolute values different from each other. 
For both non-bipartite and bipartite graphs, we propose distributed methods to compute the biases.

The problem considered in this paper can be further investigated. 
First, if the sensors are affected by noise in addition to biases, one could study how  noise 
impacts the accuracy of the estimation of the biases \cite{lee2015secure}. 
Second, the result for bipartite graphs could be also used for problems where the range and angle of arrival measurements are affected  by biases \cite{Meng2016FormationCW,bolognani2010consensus}-\nocite{Lee2016DistributedFC,Oh2014FormationCA,Giridhar2006DistributedCS}\cite{hashimoto2018distributed}. 


\setlength{\parskip}{0pt}
\bibliographystyle{IEEEtran}
\bibliography{rlstbs_consensu}

\end{document}